\documentclass[%
 reprint,
superscriptaddress,
 amsmath,
 amssymb,
 aps,
floatfix,
]{revtex4-2}

\usepackage{graphicx}%
\usepackage{dcolumn}%
\usepackage{bm}%
\usepackage[colorlinks,linkcolor=blue,anchorcolor=blue,urlcolor=blue, citecolor=blue]{hyperref}
\usepackage{float}
\usepackage{physics}
\usepackage{lipsum}  
\usepackage{mathtools}
\usepackage{subfig}
\usepackage{array}
\usepackage{diagbox}
\usepackage{caption}
\usepackage{siunitx}
\usepackage{placeins}
\usepackage{afterpage}
\usepackage[table]{xcolor}
\usepackage{etoolbox}

\newcommand{\pt}[1]{\left( #1 \right)}
\newcommand{\pq}[1]{\left[ #1 \right]}
\newcommand{\pg}[1]{\left\{#1 \right\}}

\newcommand{\wg}[1]{f_{\scriptscriptstyle {WG}}\pt{#1}}

\newcommand{\VigoAffiliations}{
\affiliation{Vigo Quantum Communication Center, University of Vigo, Vigo E-36310, Spain}
\affiliation{Escuela de Ingeniería de Telecomunicación, Department of Signal Theory and Communications, University of Vigo, Vigo E-36310, Spain}
\affiliation{AtlanTTic Research Center, University of Vigo, E-36310, Spain}
}

\begin{document}

\preprint{APS/123-QED}

\title{Characterising higher-order phase correlations in gain-switched laser sources \\ with application to quantum key distribution}

\author{Alessandro Marcomini}
\email{amarcomini@vqcc.uvigo.es}
\VigoAffiliations
\author{Guillermo Currás-Lorenzo}
\VigoAffiliations
\author{Davide Rusca}
\VigoAffiliations
\author{Angel Valle}
\affiliation{Instituto de Física de Cantabria, CSIC-{University} of Cantabria, Santander E-39005,
Spain}
\author{Kiyoshi Tamaki}
\affiliation{Faculty of Engineering, University of Toyama, Gofuku 3190, Toyama 930-8555, Japan}
\author{Marcos Curty}
\VigoAffiliations

\date{\today}%

\begin{abstract}
    Multi-photon emissions in laser sources represent a serious threat for the security of quantum key distribution (QKD). While the decoy-state technique allows to solve this problem, it requires uniform phase randomisation of the emitted pulses. However, gain-switched lasers operating at high repetition rates do not fully satisfy this requirement, as residual photons in the laser cavity introduce correlations between the phases of consecutive pulses. Here, we introduce experimental schemes to characterise the phase probability distribution of the emitted pulses, and demonstrate that an optimisation task over interferometric measures suffices in determining the impact of arbitrary order correlations, which ultimately establishes the security level of the implementation according to recent security proofs. We expect that our findings may find usages beyond QKD as well.

\end{abstract}

\maketitle

\newpage
\section{Introduction}\label{sec:Intro} 
Quantum key distribution (QKD) promises to enable information-theoretically secure communications by exploiting the laws of quantum mechanics. In particular, it allows for the establishment of a symmetric secret key between two legitimate parties (typically called Alice and Bob) in such a way that any eavesdropping attempt by an adversary (Eve) cannot go undetected \cite{loSecureQuantum2014, xuSecureQuantum2020, pirandolaAdvancesQuantum2020}.

Forty years after its initial proposal \cite{bennettQuantumCryptography1984}, QKD has grown into a mature technology which has recently seen its first networks being deployed in metropolitan and inter-urban areas \cite{dynesCambridgeQuantum2019, stuckiLongtermPerformance2011, sasakiFieldTest2011,chenIntegratedSpacetoground2021} as well as over ground-to-satellite links \cite{liaoSatellitetogroundQuantum2017, chenIntegratedSpacetoground2021}. Nevertheless, to become widely adopted it must overcome certain challenges, a major one being the difficulty in theoretically proving the security of QKD implementations with realistic devices. Importantly, when these do not meet the assumptions required by security proofs, side-channels arise, leading to unnoticed information leakages.

A prominent example of security loophole is multi-photon emissions. As of today, practical QKD implementations rely on laser sources emitting phase-randomised weak coherent pulses (PR-WCPs) due to the lack of efficient single-photon sources at telecom wavelengths. PR-WCPs can effectively be regarded as a classical mixture of photon-number states, including multi-photon states. This leaves the system vulnerable to so-called photon-number-splitting (PNS) attacks \cite{brassardLimitationsPractical2000} that severely limit the achievable secret key rate and distance \cite{gottesmanSecurityQuantum2004}. 

This vulnerability can be overcome by means of the decoy-state method \cite{hwangQuantumKey2003, wangBeatingPhotonNumberSplitting2005, loDecoyState2005, maPracticalDecoy2005, limConciseSecurity2014}. 
By emitting PR-WCPs of different intensities, Alice and Bob can tightly estimate the yield and phase-error rate of the single-photon components, thus achieving a performance comparable to that provided by ideal single-photon sources. 
Today, the decoy-state method is adopted in long-distance QKD experiments over fiber \cite{zhaoExperimentalQuantum2006, rosenbergLongDistanceDecoyState2007, liuDecoystateQuantum2010, frohlichLongdistanceQuantum2017, yuan10MbQuantum2018, boaronSecureQuantum2018} and free-space \cite{liaoSatellitetogroundQuantum2017, chenIntegratedSpacetoground2021, schmitt-manderbachExperimentalDemonstration2007, liaoSatelliteRelayedIntercontinental2018}, as well as in commercial systems \cite{HttpsWwwb, HttpsWww, HttpsWwwa}.

Importantly, decoy-state QKD {typically} requires an uniform distribution of the optical phases of the WCPs. To achieve this, one can employ phase modulators together with true random number generators to manually adjust the phase of each pulse \cite{zhaoExperimentalQuantum2007}. This introduces additional complexity and ulterior assumptions on the system components, which ultimately require a dedicated security analysis, as the number of possible phases is finite \cite{caoDiscretephaserandomizedCoherent2015, curras-lorenzoTwinFieldQuantum2021, XoelSixtoPhaseMod}. A second, more popular approach consists of gain-switching the laser source (i.e., periodically modulating {its} driving current above and below threshold) and letting each pulse grow from spontaneous emissions with inherent uniform phase distribution \cite{yuanUnconditionallySecure2007, dixonGigahertzDecoy2008, lucamariniEfficientDecoystate2013}. Nevertheless, as the repetition rate {of the source} grows to match industry requirements, the intrapulse period becomes comparable with the photon lifetime in the laser cavity and, crucially, phase correlations arise \cite{kobayashiEvaluationPhase2014, grunenfelderPerformanceSecurity2020}. This violates a crucial assumption of the decoy-state method, as the conditional probability density function (PDF) of a pulse phase given the ones correlated with it is no longer uniform.

A security proof for decoy-state QKD in presence of phase correlations has been recently proposed in \cite{curras-lorenzoSecurityQuantum2024}. It requires to determine how large the uniform component of the conditional phase distribution is, namely a parameter $q$ satisfying
\begin{equation}\label{eqn: q def}
     f\pt{\phi_i|\phi_{i-\ell_c},\ldots,\phi_{i-1},\phi_{i+1},\ldots, \phi_{i+\ell_c}} \ge \frac{q}{2\pi} ,
\end{equation}
for each modulation period $i$, where 
$\ell_c$ is the (finite) range of correlations.
{Such parameter satisfies} $q \in \left[0,1\right]$, being $q=1$ the optimal case (i.e., the perfect phase-randomisation scenario). Remarkably, while \cite{curras-lorenzoSecurityQuantum2024} enables high-rate decoy-state QKD with gain-switched lasers, its application to real setups remains unfeasible as no recipe for the estimation of the parameter $q$ exists for $\ell_c > 1$.

{In this paper we address this crucial problem and provide an experimental scheme to characterise $q$ for phase correlations of arbitrary length $\ell_c$. Our scheme can be applied to test the source both at the beginning of the protocol and during its execution, as it does not affect the state preparation. In detail, we first focus on developing a model of the conditional phase distribution at the laser output on the basis of fundamental studies about stochastic processes in the laser cavity when residual photons from previous pulses are present. Afterwards, we extend the known analysis for determining $q$ in the case of nearest neighbour correlations to arbitrary correlations lengths, by designing efficient experimental schemes based on interferometric measurements. Importantly, we show how these suffice in quantifying the relative strength of spontaneous emissions compared to the stimulated ones, which our analysis suggests being the critical parameter for security. Remarkably, our proposal allows to estimate all the relevant parameters simply through an optimisation task over amplitude attenuators and phase shifters in the interferometer.

The paper is organised as follows. In Sec. \ref{sec:assumptions} we state under which assumptions our analysis holds and we provide the general description of our model. In Sec. \ref{sec:FirstOrder} we recap some known results for first-order correlations as a basis for our study, while in Sec. \ref{sec:SecondOrder} we depict the natural generalisation to the second-order case. Here we also introduce a three-line interferometer that serves the purpose of estimating the phase uncertainty in this regime and we evaluate numerically the dependence of $q$ on experimental data. Next, in Sec. \ref{sec:HigherOrder} we describe the general scenario of correlations of arbitrary order,
while in Sec. \ref{sec:sim} we run a virtual experiment and infer the security level of a laser in different operational regimes by adopting our scheme.
Finally, Sec. \ref{sec: Conclusions} provides a summary of our results. Additional calculations and considerations on the phase model are reported in the Appendices.}

\section{Assumptions}\label{sec:assumptions}
The sequence of global phases $\pmb{\phi}_1,\ldots,\pmb{\phi}_N$ of the WCPs emitted by Alice constitutes a discrete-time stochastic process with associated joint PDF
\begin{equation}
    f\pt{\pmb{\phi}_1=\phi_1,\ldots,\pmb{\phi}_N=\phi_n} \equiv f\pt{\phi_1,\ldots,\phi_N} .
\end{equation}
The analysis we propose relies on the following assumptions:
\begin{itemize}
    \item[(A1)] At each round Alice's laser output is a single-mode coherent state of constant intensity.
    \item[(A2)] The stochastic process is \textit{generalised-Markovian}, that is, it holds at most $\ell_c$ rounds of memory with $\ell_c$ finite and known. This means that
    \begin{equation}
        f\pt{\phi_i|\phi_{i-1},\ldots,\phi_{1}} = f\pt{\phi_i| \phi_{i-1},\ldots,\phi_{i-\ell_c}} \ \forall i.
    \end{equation}
    This is a direct requirement of the security proof introduced in \cite{curras-lorenzoSecurityQuantum2024}. Note that in Appendix \ref{sec: exponential decay} we lift this assumption while addressing the specific case of exponentially decaying correlations.
    \item[(A3)] The process is conditionally stationary, i.e.:
    \begin{align}
        &f\pt{\pmb{\phi}_i=\phi_i|\pg{\pmb{\phi}_j=\phi_j, j = i-1, \ldots, i-\ell_c}} \nonumber \\
        &= f\pt{\pmb{\phi}_{i+k}=\phi_i|\pg{\pmb{\phi}_{j+k}=\phi_j, j = i-1, \ldots, i-\ell_c}} ,
    \end{align}
    $\forall k \in \mathbb{N}.$
    That is, the analytical shape of the conditional PDF (together with its parameters) does not depend on the round index.
    \item[(A4)] The phase of the $i-$th round can be treated as a random variable defined by
    \begin{equation}\label{eqn: introcudion of phi hat and delta phi}
        \pmb{\phi}_i = \hat{\pmb{\phi}}_i^{\pt{\ell_c}} + \pmb{\delta\phi}_i \mod 2\pi,
    \end{equation}
    where
    \begin{align}
    \hat{\pmb{\phi}}_i^{\pt{\ell_c}} &:=h\pt{\pmb{\phi}_{i-1}, \ldots, \pmb{\phi}_{i-\ell_c}} \nonumber \\
        &=\arg \pt{\sum_{n=i-\ell_c}^{i-1} r_{n}e^{j\pmb{\phi}_n}} \label{eqn: phi hat generic} ,
    \end{align}
    for some (possibly unknown) parameters $\pg{r_n}_n$. Here $\pg{\pmb{\delta\phi}_i}_i$ is a sequence of independent and identically distributed (iid) Gaussian random variables, which effectively represents the spontaneous emission noise between pulses and is responsible for phase uncertainty. The physical meaning of the terms $\pg{r_n}_n$ is that of the relative amount of photons from round $n$ remaining is the cavity, and Eq.~\eqref{eqn: phi hat generic} indicates that residual photons in the cavity combine in a constructive way \cite{glauberCoherentIncoherent1963}.
    Note that Assumption (A3) ensures that the terms $\pg{r_n}_n$ are the same for every round and, since the function $\arg\pt{x}$ is invariant for rescaling of the input $x$, we can select them such that $r_{i-1}=1$. This assumption generalises the known results for the simplest case $\ell_c = 1$ \cite{curras-lorenzoSecurityQuantum2024, kobayashiEvaluationPhase2014}.
\end{itemize}

In Appendix \ref{app: proof of PDF assumption} we prove that Assumption (A4) directly implies that the conditional PDF of a phase given the previous ones is a wrapped Gaussian (WG) distribution (see definition in Eq.~\eqref{eqn: wrapped gaussian definition}), that is
\begin{align}
    &f( \phi_i|  \phi_{i-1},\ldots,\phi_{i-\ell_c}) =\wg{\phi_i;\hat{\phi}_{i}^{\pt{\ell_c}} + \overline{\delta\phi},\sigma_{\ell_c}},\label{eqn: A4 assumption model A}
\end{align}
where $\hat{\phi}_i^{\pt{\ell_c}} := h\pt{\phi_{i-1}, \ldots, \phi_{i-\ell_c}}$, while
$\overline{\delta\phi}$ and $\sigma_{\ell_c}$ are the central value and standard deviation of the PDF of the random variables $\pg{\pmb{\delta\phi}_i}_i$. 
Importantly, this PDF can be equivalently considered being a wrapped Gaussian itself, as $\pmb{\delta\phi}_i$ enters Eq.~\eqref{eqn: introcudion of phi hat and delta phi} modulo $2\pi$. 

A supplementary note on the physical reasoning behind the formulation of this model is reported in Appendix \ref{sec: Note on the cavity model}, where we further discuss how the past phase realisations (namely, the values $\phi_{i-1},\ldots,\phi_{i-\ell_c}$) determine the most likely phase at round $i$, while the uncertainty is completely given by the strength of spontaneous emissions.

\begin{figure}
    \centering
    \includegraphics[width=\columnwidth]{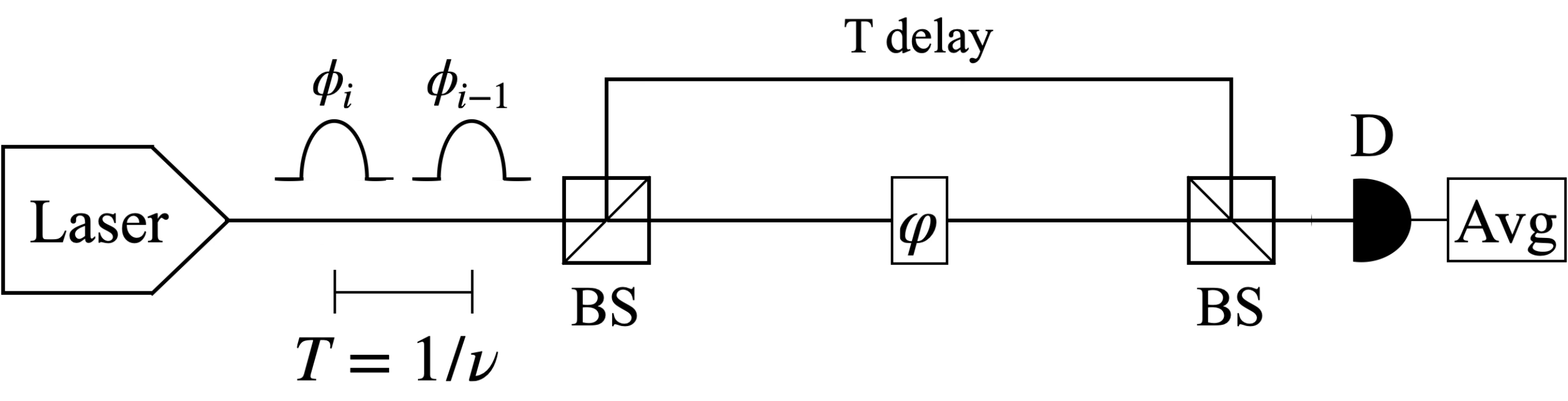}
    \caption{Experimental scheme of the asymmetric Mach-Zender interferometer for the visibility measure of first order correlations. A delay line of length $T=1/\nu$, being $\nu$ the repetition rate of the source, is used to make nearest neighbouring pulses of strong light interfere. Here $\varphi$, phase shifter; D, {linear detector}; Avg, averager. Beam splitters (BSs) are considered $50:50$.}
    \label{fig:KobTomExp}
\end{figure}

\section{\label{sec:FirstOrder} First order correlations}
Here we report the analysis originally developed in \cite{curras-lorenzoSecurityQuantum2024} for the case $\ell_c=1$, which serves as a basis for further generalisation in the next section.

By the law of total probability, Eq.~\eqref{eqn: q def} becomes
\begin{align}
     \frac{q}{2\pi}
     &= \min_{\phi_{i-1},\phi_{i},\phi_{i+1}} \frac{f\pt{\phi_{i+1},\phi_{i},\phi_{i-1}}}
     {f\pt{\phi_{i+1},\phi_{i-1}}} \nonumber \\
     &= \min_{\phi_{i-1},\phi_{i},\phi_{i+1}} \frac{f\pt{\phi_{i+1},\phi_{i},\phi_{i-1}}}
     {\int_{-\pi}^{\pi}f\pt{\phi_{i+1},{\phi}'_i,\phi_{i-1}}\dd{\phi}'_i} \nonumber \\
     &= \min_{\phi_{i-1},\phi_{i},\phi_{i+1}} \frac{f\pt{\phi_{i+1}|\phi_{i}}f\pt{\phi_i|\phi_{i-1}}}
     {\int_{-\pi}^{\pi} f\pt{\phi_{i+1}|\phi'_{i}}f\pt{\phi'_i|\phi_{i-1}} \dd{\phi}'_i}, \label{eqn: full q lc 1}
\end{align}
which can be numerically solved once an analytical form for the conditional PDF $f\pt{\phi_i|\phi_{i-1}}$ is found.

From Eqs.~\eqref{eqn: phi hat generic}-\eqref{eqn: A4 assumption model A}, we have that the conditional PDF at each modulation period is a WG distribution centered around the previous phase, up to a constant shift $\overline{\delta\phi}$. In fact, we have 
\begin{equation}
    \hat{\pmb{\phi}}_i^{\pt{1}} = \arg\pt{r_{i-1}e^{j\pmb{\phi}_{i-1}}} = \pmb{\phi}_{i-1}.
\end{equation}
Therefore, to numerically find $q$ following Eq.~\eqref{eqn: full q lc 1} one only needs to estimate $\overline{\delta\phi}$ and $\sigma_1$.

As proposed in \cite{kobayashiEvaluationPhase2014}, the latter can be found by measuring the visibility $\mathcal{V}$ of the interference of neighbour pulses in an asymmetric, balanced Mach-Zender interferometer like that shown in Fig.~\ref{fig:KobTomExp}. In fact, it can be demonstrated that \cite{kobayashiEvaluationPhase2014}:
\begin{equation}\label{eqn: visibility definition}
    \mathcal{V} = \frac{I\pt{\varphi_{max}} - I\pt{\varphi_{min}}}{I\pt{\varphi_{max}}+I\pt{\varphi_{min}}} = \exp(-\frac{\sigma_1^2}{2}) ,
\end{equation}
where $I\pt{\varphi}$ is the average intensity of the interfered light found by tuning the phase $\varphi$ (see Fig.~\ref{fig:KobTomExp}). Here, $\varphi_{max}$ ($\varphi_{min}$) represents the value of the phase shifter returning the maximum (minimum) intensity signal, which occurs for $\varphi_{max} = -\overline{\delta\phi}$ ($\varphi_{min} = \pi -\overline{\delta\phi}$).
Combined with Eq.~\eqref{eqn: A4 assumption model A}, Eq.~\eqref{eqn: full q lc 1} yields \cite{curras-lorenzoSecurityQuantum2024}:
\begin{equation}
    \frac{q}{2\pi} = \min_{\phi_{i-1}',\phi_{i},\phi_{i+1}''} \frac{\wg{\phi_{i+1}'';\phi_{i},\sigma_1}\wg{\phi_i;\phi_{i-1}',\sigma_1}}
     {\wg{\phi_{i+1}'';\phi_{i-1}',\sqrt{2}\sigma_1}} ,
\end{equation}
where $\phi_{i-1}' = \phi_{i-1} + \overline{\delta\phi}$, $\phi_{i+1}'' = \phi_{i+1} - \overline{\delta\phi}$ and
\begin{equation}\label{eqn: sigma_1 in terms of V}
    \sigma_1 = \sqrt{-2\ln(\mathcal{V})} .
\end{equation}

Remarkably, a recent experiment at an emission rate of $5$ GHz returned a visibility $\mathcal{V} = 0.0019$ \cite{grunenfelderPerformanceSecurity2020}, corresponding to a standard deviation $\sigma_1 = 3.54003$ rad and a value of the parameter $q$ equal to $q = 0.9924$ \cite{curras-lorenzoSecurityQuantum2024}.

\begin{figure}
    \centering
    \includegraphics[width=\columnwidth]{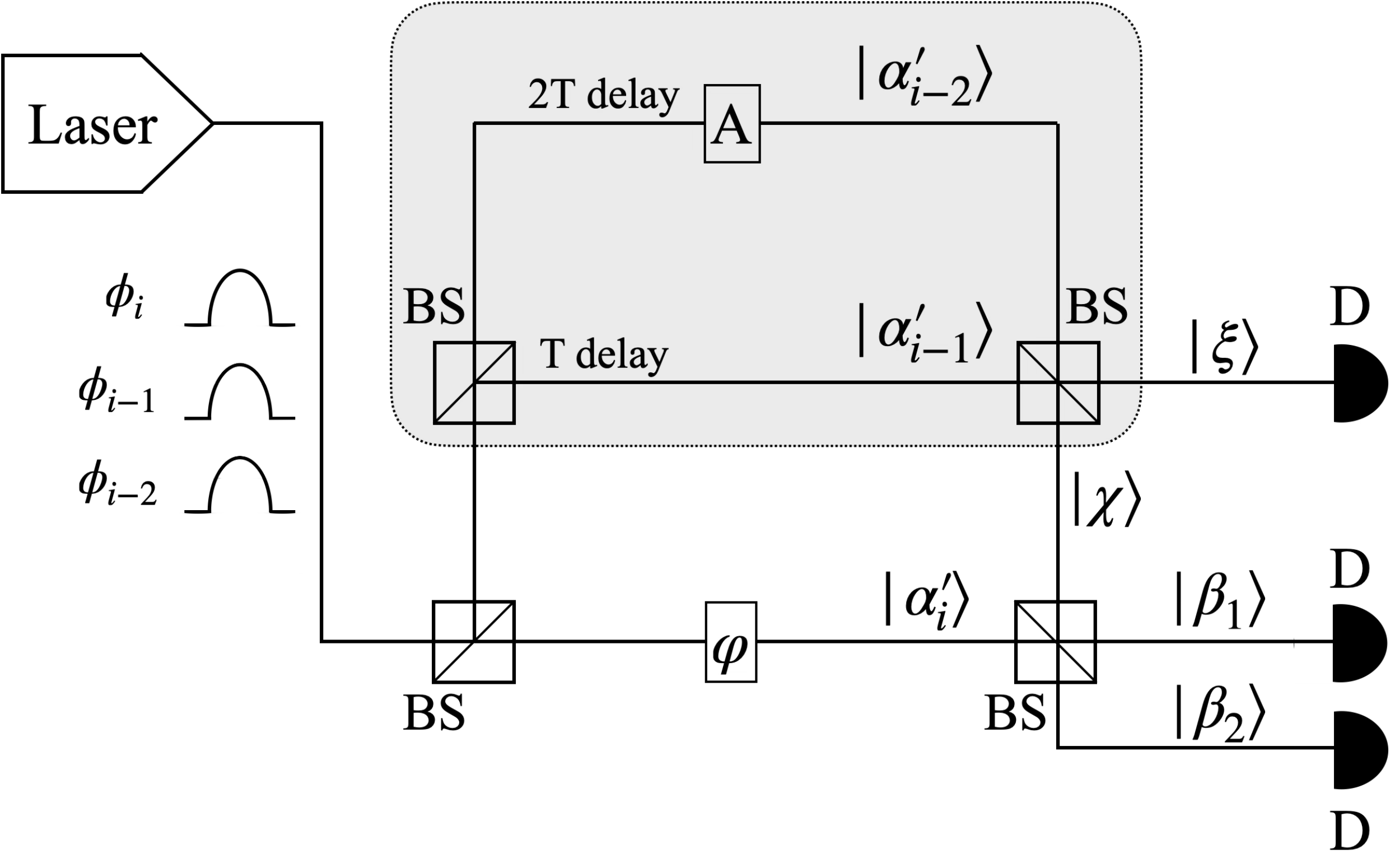}
    \caption{Experimental scheme for the estimation of correlation strength up to the second order. The interferometer module in the grey box takes the role of replicating the way light combines during the buildup process in the laser cavity.
    When the ratio on the intensities of the states $\ket{\alpha'_{i-2}}$ and $\ket{\alpha'_{i-1}}$ matches {$r_{i-2}$,   the state $\ket{\chi}$ has phase $\hat{\pmb{\phi}}^{\pt{2}}_{i}$}. Therefore, measuring the visibility of the interference of this state with $\ket{\alpha_i}$ allows to quantify the quantum noise due to spontaneous emissions.
    Here, {$\varphi$, phase shifter; D, linear detector; A, attenuator. Beam splitters (BSs) are considered $50:50$.}}
    \label{fig:exp_setup}
\end{figure}

\section{\label{sec:SecondOrder} Second order correlations}

Let us now focus on the case $\ell_c = 2$. Following a similar reasoning to Sec. \ref{sec:FirstOrder}, one can easily prove that the generalisation of Eq.~\eqref{eqn: full q lc 1} to this scenario is
\begin{equation}\label{eqn: general q formula 2}
    \frac{q}{2\pi} = \min_{\pg{\phi_n}_{n=i-2}^{i+2}} \frac{\prod_{k=0}^{2}f\pt{\phi_{i+k}|\phi_{i+k-1},\phi_{i+k-2}}}{f\pt{\phi_{i+2},\phi_{i+1}|\phi_{i-1},\phi_{i-2}}} ,
\end{equation}
where
\begin{align} \label{eqn: denom for ell_c=2}
    f\pt{\phi_{i+2},\phi_{i+1}|\phi_{i-1},\phi_{i-2}}
    =\int_{-\pi}^{\pi} &f\pt{\phi_{i+2}|\phi_{i+1},\phi'_i}\\
     \times &f\pt{\phi_{i+1}|\phi'_i,\phi_{i-1}} \nonumber\\
     \times &f\pt{\phi'_i|\phi_{i-1},\phi_{i-2}}\dd{\phi}'_i .
     \nonumber
\end{align}

Note that by substituting Eq.~\eqref{eqn: denom for ell_c=2} in Eq.~\eqref{eqn: general q formula 2}, all terms involved are in the form given by Eq.~\eqref{eqn: A4 assumption model A}.
From the latter, we have that the conditional PDF of a phase given previous ones is characterised by the central value and standard deviation of the distribution $\pmb{\delta\phi}_i$, together with the term $r_{i-2}$ in Eq.~\eqref{eqn: phi hat generic}. Crucially, we note that by reproducing pulses with phase $ \pmb{\delta\phi}_i$, the same approach of section \ref{sec:FirstOrder} would allow to estimate through interference visibility the parameters $\overline{\delta\phi}$ and $\sigma_2$, which can later be used to numerically optimise Eq.~\eqref{eqn: general q formula 2}.
Therefore, the key idea here is to adopt the scheme illustrated in the grey box in Fig.~\ref{fig:exp_setup} to replicate the coherent state combination that occurs inside the cavity and to produce a state $\ket{\chi}$ having phase $\hat{\pmb{\phi}}_i^{\pt{2}}$. As a result, the output state $\ket{\beta_2}$ in Fig.~\ref{fig:exp_setup} has phase
\begin{equation}
     \pmb{\delta\phi}_i = \pmb{\phi}_i - \hat{\pmb{\phi}}_i^{\pt{2}}.
\end{equation}

In detail, consider a laser emitting WCPs at a repetition rate $\nu = 1/T$. These undergo a series of beam splitters {(BSs)} that guide them into the three channels of a cascade asymmetric Mach-Zender interferometer, which are 
tuned in the optical path length so to provide no delay, $T$ and $2T$ delay, respectively. 
At round $i$, a train of pulses having phases $\pmb{\phi}_i,\ \pmb{\phi}_{i-1}$ and $\pmb{\phi}_{i-2}$ produced at times $t_i > t_{i-1} > t_{i-2}$ goes through the channels. The pulse amplitudes entering the channels are determined by the transmittance of the BSs on the input side. Through this paper we always consider $50:50$ lossless BSs for simplicity, but one might prefer other settings to balance the amount of light in each arm. 

In our setup we accommodate for an amplitude attenuator $A$ and a phase shifter $\varphi$ such that the coherent states in Fig.~\ref{fig:exp_setup} take the {form
\begin{align} \nonumber
    \ket{\alpha'_{i-2}} = \ket{\sqrt{\mu'_{i-2}}e^{j\pmb{\phi}_{i-2}}}, \\
    \ket{\alpha'_{i-1}} = \ket{\sqrt{\mu'_{i-1}}e^{j\pmb{\phi}_{i-1}}}, \label{eqn: alpha prime states main text}\\
    \ket{\alpha'_{i}} = \ket{\sqrt{\mu'_i}e^{j\pt{\pmb{\phi}_i + \varphi}}} . \nonumber
\end{align}}
Here,
\begin{equation}\label{eqn: relation A to mu'}
    {\mu'_{i-2}} = {A\ \frac{\mu_{i-2}}{4}}, \quad {\mu'_{i-1}} = {\frac{\mu_{i-1}}{4}}, \quad {\mu'_i} = {\frac{\mu_i}{2}},
\end{equation}
{where $\mu_{n}$ for $n\in\pg{i-2,i-1,i}$ denotes the intensity of the $n-$th pulse at the entrance of the interferometer.
By tuning the value of $A$ in the longest delay line} it is possible to change the relative amplitude of the state $\ket{\alpha'_{i-2}}$ with respect to $\ket{\alpha'_{i-1}}$, that is 
\begin{equation}
    r'_{i-2} := \sqrt{\mu'_{i-2}/\mu'_{i-1}}.
\end{equation}
Importantly, as further discussed in Appendix \ref{AppSec: g}, only slowly tunable attenuation is required, i.e., to be adjustable over a large number of rounds.

If an either direct (e.g., by measuring field components in the cavity) or indirect (via laser rate equations) reliable estimate of the value of the term $r_{i-2}$ in Eq.~\eqref{eqn: phi hat generic} is accessible, then it would be sufficient to set the attenuator $A$ such that $r'_{i-2} = r_{i-2}$. In this way, the phase of the state $\ket{\chi}$ in Fig.~\ref{fig:exp_setup} matches $\hat{\pmb{\phi}}_i^{\pt{2}}$.
On the other hand, if the value of $r_{i-2}$ is unknown, in Appendix \ref{AppSec: g} we prove that one can find its most accurate estimation by maximising over the possible settings of $\varphi$ and $A$ the quantity
\begin{equation}\label{eqn: definitioon of g}
    v^{\pt{2}} := \expval{\frac{{2\pmb{\mu}_{\beta_1} +\pmb{\mu}_{\xi}} - \sum_{n=i-2}^i\mu'_n}{2\sqrt{\mu'_i\pt{\mu'_{i-1} + \mu'_{i-2} - \pmb{\mu}_{\xi}}}}} .
\end{equation}
Here $\pmb{\mu}_{\beta_1}$ ($\pmb{\mu}_{\xi}$) is the intensity of the {state $\ket{\beta_1}$ ($\ket{\xi}$)} measured by the linear detectors. As they depend on the values taken by the random variables $\pmb{\phi}_{i-2}$, $\pmb{\phi}_{i-1}$ and $\pmb{\phi}_{i}$, they are actually random variables themselves.
{Denoting by $\varphi_{max}$ and $A_{max}$ the values of the phase shifter and attenuator maximising Eq.~\eqref{eqn: definitioon of g}, we find that the best estimate of $r_{i-2}$ is given by
\begin{equation}\label{eqn: optimal paramteres main text 1} 
    r_{i-2} =  \sqrt{A_{max}\frac{\mu_{i-2}}{\mu_{i-1}}}.
\end{equation}
Moreover, we find that similarly to the case $\ell_c=1$ it holds \begin{equation}\label{eqn: optimal paramteres main text 2} 
     \overline{\delta\phi} = - \varphi_{max}.
\end{equation}}

\begin{figure*}
\hfill
\subfloat[$q$ vs $\sigma_2$ for fixed $r_{i-2}$.]{\includegraphics[width=\columnwidth]{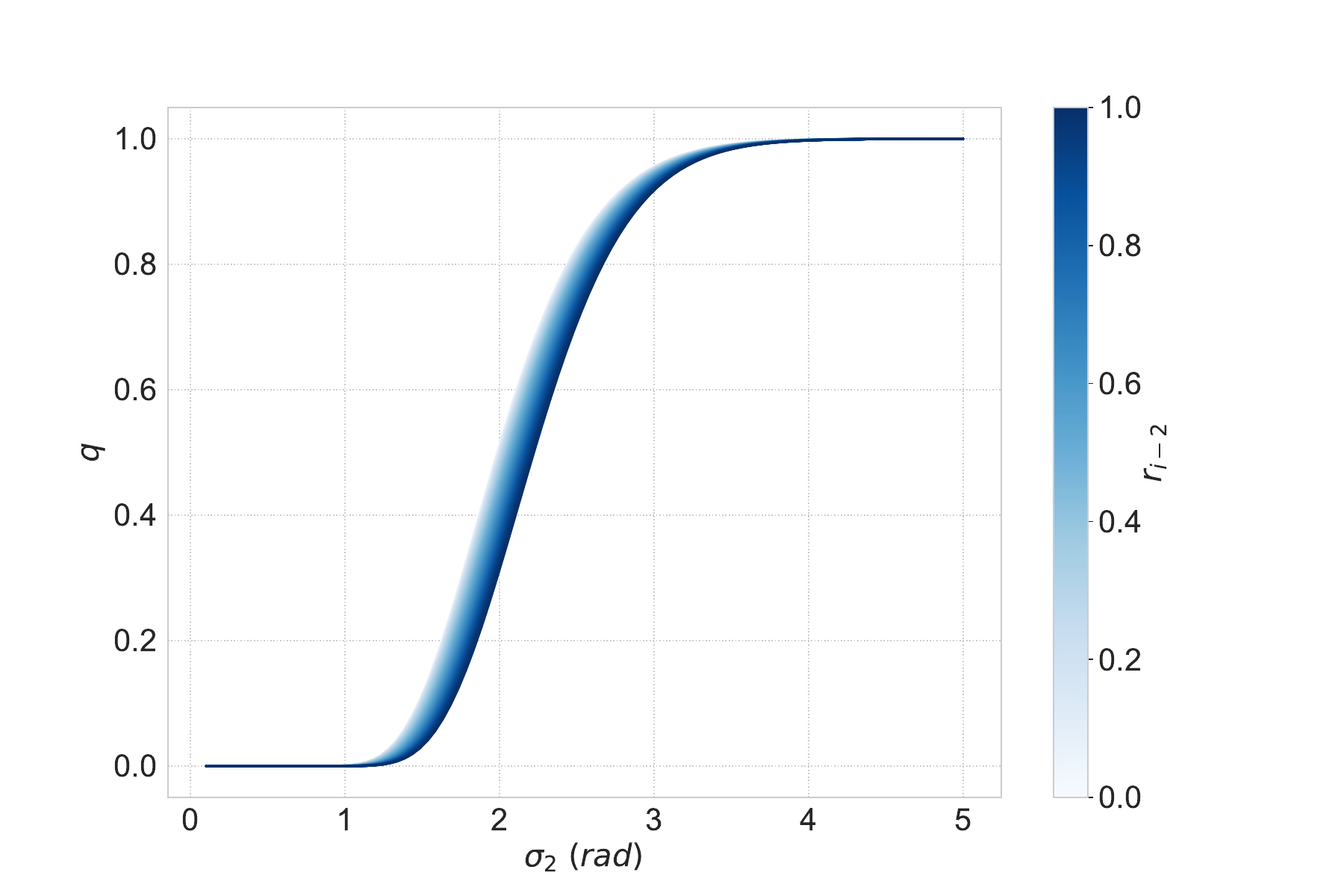}}\hfill
\subfloat[$q$ vs $r_{i-2}$ for fixed $\sigma_2$.]{\includegraphics[width=\columnwidth]{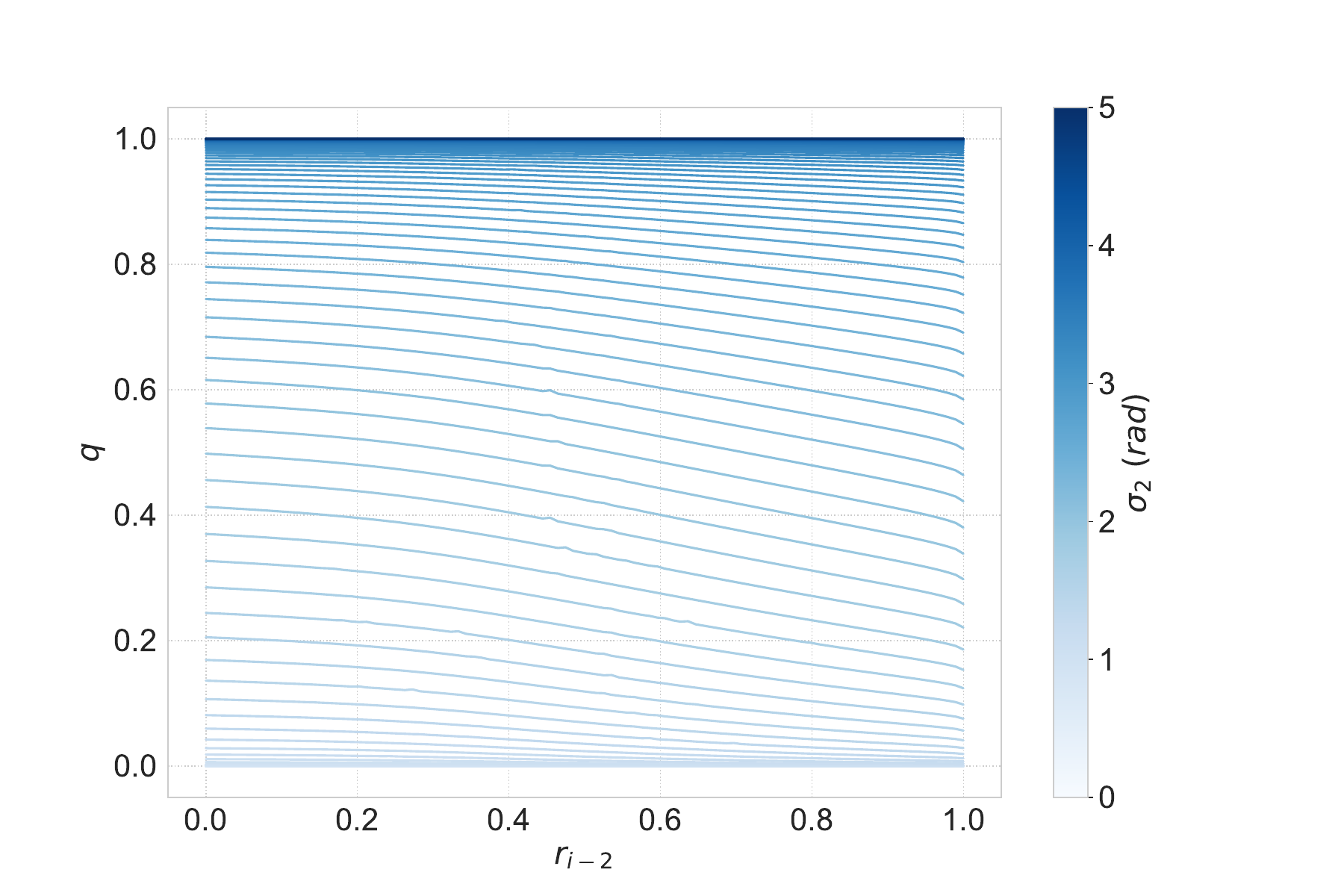}}    
\hfill
\caption{Results for the numerical minimisation of Eq.~\eqref{eqn: general q formula 2} with $\overline{\delta\phi}=0$. The value of $q$ for second order correlations is computed numerically upon changing the values of $\sigma_2$ (subfigure (a)) and $r_{i-2}$ (subfigure (b)) in the definition of the wrapped Gaussian distribution.}
\label{fig: q vs R and s}%
\end{figure*}

We note that the expectation value in Eq.~\eqref{eqn: definitioon of g} corresponds to the ensemble average but crucially, since the random variables $\pg{\pmb{\delta\phi}_i}_i$ are iid, the stochastic process underlying is ergodic. As a consequence, one can find a reliable estimate of the ensemble average by taking the mean value of realisations over a large number of rounds, which is easily implementable in real setups.
Equations involving average quantities in this paper must be read keeping this in mind.

In particular, one can estimate $\sigma_2$ as:
\begin{equation}\label{eqn: estimate for sigma_2 in expscheme}
    \sigma_2 = \sqrt{-2\ln\pt{\max_{\varphi, A} v^{\pt{2}}}} .
\end{equation}
Note that direct comparison with Eq.~\eqref{eqn: sigma_1 in terms of V} suggests that $v^{\pt{2}}$ takes the role of a generalised visibility measure for $\ell_c=2$ (more comments on this in Appendix \ref{AppSec: g}).

In short, the experimental setup in Fig.~\ref{fig:exp_setup} allows for a direct measurement of the amount of spontaneous emissions in the cavity during the lasing process (and therefore the level of phase randomness) simply by sweeping over a channel attenuation and a local phase shifter, and monitoring the output. 

Given {the estimates of $r_{i-2}$ and $\overline{\delta\phi}$ in Eqs.~\eqref{eqn: optimal paramteres main text 1}-\eqref{eqn: optimal paramteres main text 2}, the value of $q$ can finally be found by solving Eq.~\eqref{eqn: general q formula 2} numerically. This step is further investigated in the next section.}

\subsection*{Numerical estimation of q}

In the previous section we introduced a method to experimentally estimate the values of {$r_{i-2}$, $\sigma_2$ and $\overline{\delta\phi}$}. Here we investigate how the value of $q$ is affected by {the first two,} focusing on the case $\overline{\delta\phi}=0$ for simplicity. Results in Fig.~\ref{fig: q vs R and s} report the numerical minimisation of Eq.~\eqref{eqn: general q formula 2}
for various values of $r_{i-2}$ and $\sigma_2$. The conditional probability distribution follows the definitions in Eqs.~\eqref{eqn: phi hat generic}-\eqref{eqn: A4 assumption model A} for $\ell_c=2$.

Importantly, Fig.~\ref{fig: q vs R and s} confirms that the variance of the distribution is the critical parameter for security, as for large enough values of $\sigma_2$ the PDF given in Eq.~\eqref{eqn: A4 assumption model A} approaches a uniform profile ($q=1$), regardless of the value of $r_{i-2}$ (see Fig.~\ref{fig: q vs R and s}(a)). Dependence on the $r_{i-2}$ parameter is weaker and simulations suggest that the minimum value of $q$ for fixed $\sigma_2$ occurs for $r_{i-2}\to1$ (see Fig.~\ref{fig: q vs R and s}(b)). We believe this being due to the fact that for a given value of $\phi_{i-1}$, the central value of the $i-$th phase distribution in Eq.~\eqref{eqn: phi hat generic} is bounded by
\begin{equation}
    \phi_{i-1} - \arcsin\pt{r_{i-2}} \le \hat{\phi}^{\pt{2}}_i \le \phi_{i-1} + \arcsin\pt{r_{i-2}}.
\end{equation}
Since $\arcsin(r_{i-2})$ is monotone increasing for $r_{i-2}\in(0,1)$, a larger value of $r_{i-2}$ effectively corresponds to a larger parameter space over which the minimisation of Eq.~\eqref{eqn: general q formula 2} is performed. {In a realistic scenario it may be possible that the maximum value of Eq.~\eqref{eqn: definitioon of g} (and therefore $\sigma_2$) can be established or bounded with high confidence, while the uncertainty in the estimate of $r_{i-2}$ might be large. For example, this could happen for very weak second order effects, as when $r_{i-2}\ll1$ a change in the attenuator has no noticeable effects on $v^{\pt{2}}$. In case that a unique value of $r_{i-2}$ cannot be retrieved from the optimal settings, Fig.~\ref{fig: q vs R and s} suggests that evaluating $q$ at $r_{i-2} = 1$ for a given estimate of $\sigma_2$ can provide a good lower bound.}

Finally, we remark that in these simulations we only investigate values of $r_{i-2}$ within the range $\pt{0,1}$, as it is physically meaningful that correlations among neighbour pulses are not weaker than those  among pulses two periods apart. Still, we remark that the analysis presented is valid for any value of $r_{i-2}$.

\section{\label{sec:HigherOrder} Arbitrary order correlations}
Extension to arbitrary length of correlation $\ell_c$ follows naturally from the results of the previous section.

\begin{figure}
    \centering
    \includegraphics[width=\columnwidth]{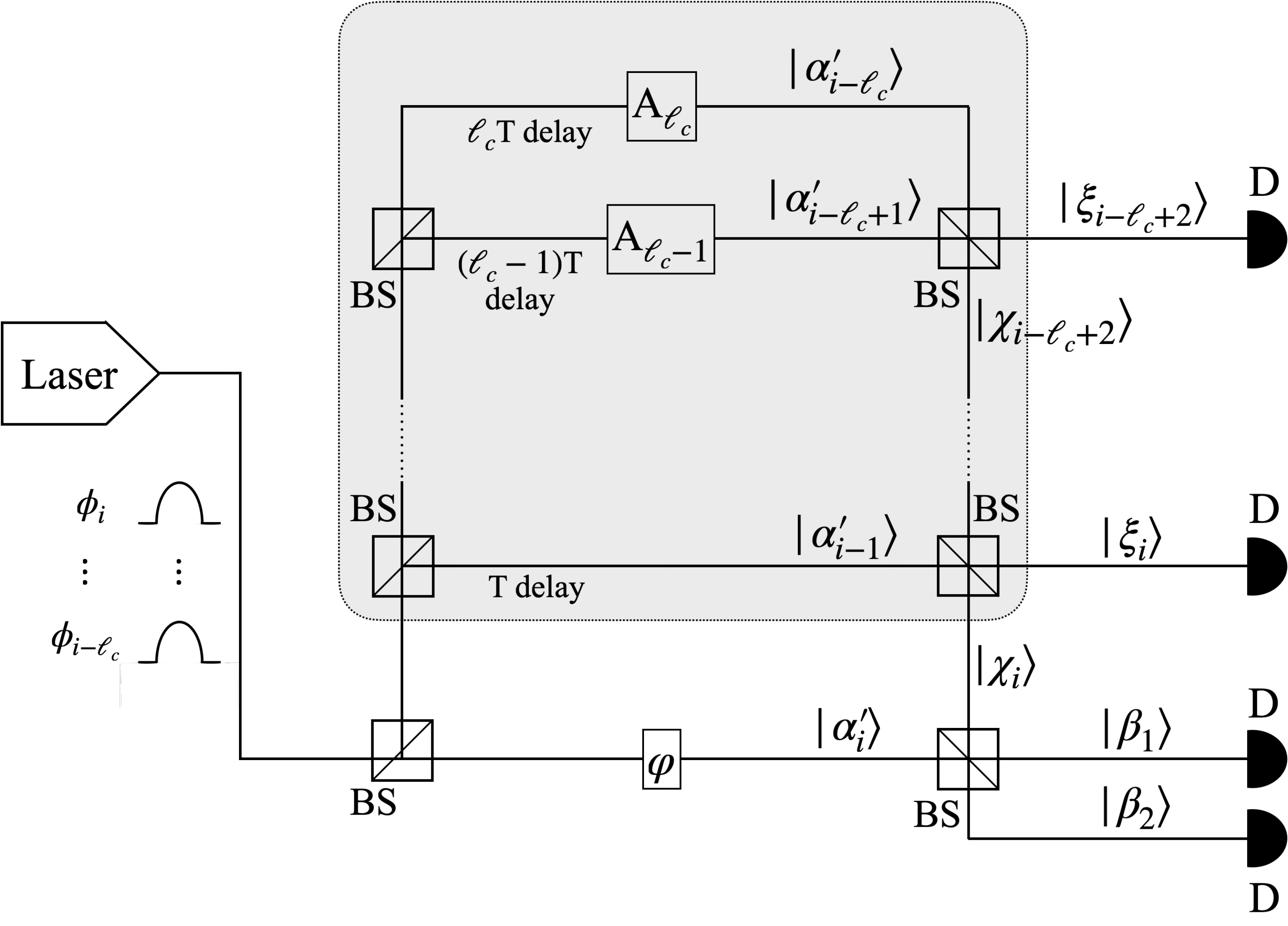}
    \caption{General scheme for the estimation of correlation strength up to $\ell_c-$th order. A congruent amount of delay lines is introduced with respect to the setup of Fig.~\ref{fig:exp_setup} to allow pulses of $\ell_c$ subsequent rounds to interfere in cascade. 
    Here, {$\varphi$, phase shifter; D, linear detector; $A_k$, attenuators. Beam splitters (BSs) are considered $50:50$.}}
    \label{fig:exp_setup_inf}
\end{figure}

Consider the scheme in Fig.~\ref{fig:exp_setup_inf}.
The introduction of additional delay lines with tunable attenuators can ensure that the state interfering with $\ket{\alpha'_i}$ at the last beam splitter has a phase $\hat{\pmb{\phi}}_i^{\pt{\ell_c}}$, upon finding the best relative intensities in the channels. 
While this might be experimentally challenging for large correlation lengths, it still provides an exact way to access the variance of the phase distribution without the need to further characterise the laser source.

The generalisation of Eq.~\eqref{eqn: definitioon of g} to arbitrary correlation length is given by
\begin{align}\label{eqn: definitioon of g for ell_c}
v^{\pt{\ell_c}} =  \expval{\frac{2 \pmb{\mu}_{\beta_1} + \sum_{k=i-\ell_c+2}^{i}\pmb{\mu}_{\xi_k} -\sum_{n=i-\ell_c}^{i}\mu'_n}{2\sqrt{\mu'_i\pt{\sum_{n=i-\ell_c}^{i-1}\mu'_n-\sum_{k=i-\ell_c+2}^{i}\pmb{\mu}_{\xi_k}}}}}.
\end{align}
Here $\pg{\pmb{\mu}_{\xi_k}}_{k=i-\ell_c+2}^{i}$ is the set of intensities of the states $\pg{\ket{\xi_k}}_k$ in Fig.~\ref{fig:exp_setup_inf}, while $\pg{\mu'_{i-\ell_c},\ldots,\mu'_i}$ the set of tunable intensities in the interferometer arms. The full derivation of Eq.~\eqref{eqn: definitioon of g for ell_c} is provided in Appendix \ref{Appendix: generalisation of g}.

{Again, one finds that optimising over $\varphi$ and $\pg{A_k}_{k=2}^{\ell_c}$:
\begin{equation}
\begin{gathered}
    \sigma_{\ell_c} = \sqrt{-2\ln\pt{\max_{\varphi ,\pg{A_k}_{k}}{v^{\pt{\ell_c}}}}},
    \\
    r_{i-k} =  \sqrt{A_{max,k}\frac{\mu_{i-k}}{\mu_{i-1}}} \quad \text{for } \quad 2\le k \le \ell_c,
\end{gathered}
\end{equation}
being $\varphi_{max}$ and $\pg{A_{max,k}}_k$ the optimal settings. The quantity $\overline{\delta\phi}$ is found though Eq.~\eqref{eqn: optimal paramteres main text 2}.
Using these estimates of $\sigma_{\ell_c}$, $\pg{r_{i-k}}_k$ and $\overline{\delta\phi}$, the value of $q$ can} be found numerically by applying the law of total probability to Eq.~\eqref{eqn: q def} and employing the definitions in Eqs.~\eqref{eqn: phi hat generic}-\eqref{eqn: A4 assumption model A}.

These results do not require any model of the way the amount of residual photons dynamically changes in the laser cavity. Nevertheless, additional knowledge about the scaling of correlations over rounds can help in the design of more efficient experimental schemes. A particularly interesting case is the one of exponential decay of correlations \cite{agulleiroModelingCharacterization2024}. We discuss a simpler experimental setup for this scenario in Appendix \ref{sec: exponential decay}.

\section{Simulation}\label{sec:sim}
In this section we run a numerical experiment with the setup in Fig.~\ref{fig:exp_setup}. To do so, we simulate the dynamics of a gain-switched laser and apply the analysis introduced in section \ref{sec:SecondOrder} to estimate $r_{i-2}$, $\overline{\delta\phi}$ and $\sigma_2$ for the case $\ell_c=2$. Detailed information on the signal generation process and the parameters used for the virtual laser are provided in Appendix \ref{appsec: LaserSim}.

For each combination of the repetition rate $\nu$ and lower driving current $I_{off}$ {of the laser,} we compute the value of $v^{\pt{2}}$ averaging over a total of $10^4$ pulses. While this value is chosen for numerical convenience, note that an experimental estimate could (and should) average over {a greater number of pulses}, so that the sample average properly converges to the ensemble average.
We compute $v^{\pt{2}}$ for different values of the phase shift $\varphi$ and attenuator $A$, in a configuration such that {$\mu_{i-2} = \mu_{i-1}$, which implies} $r'_{i-2} = \sqrt{\mu'_{i-2}/\mu'_{i-1}} = \sqrt{A}$. 

\begin{figure*}
    \centering
    \includegraphics[width=.85\textwidth]{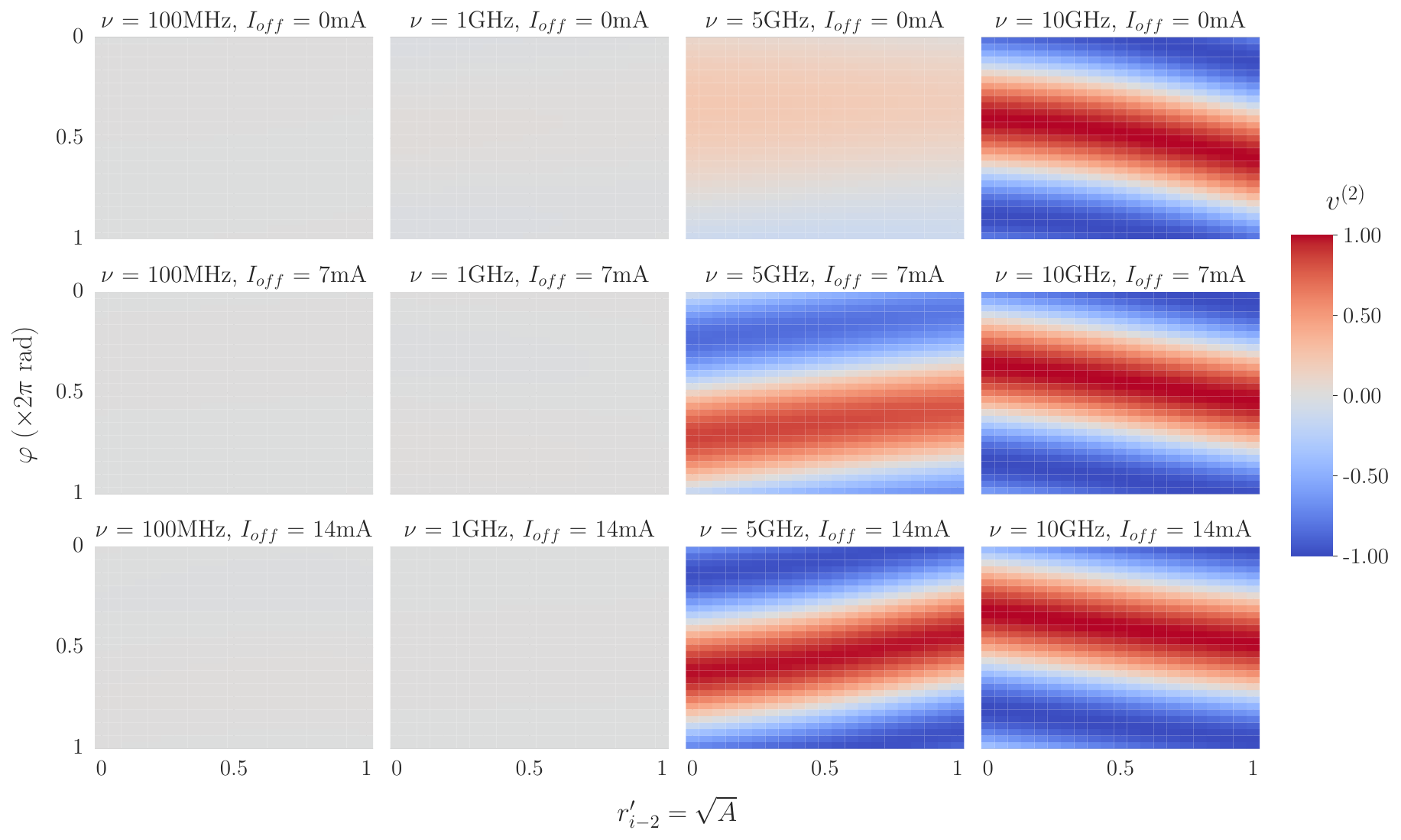}
    \caption{Average value $v^{\pt{2}}$ over $10^4$ virtual pulses in different operational regimes of the laser, upon changing the values of $\varphi$ and the attenuator $A$. By increasing the source emission rate and the value of $I_{off}$, the second order visibility is noticeably higher, implying less randomness in the pulses phases. Results are summarised in Table \ref{tab: q vals sim}.}
    \label{fig:opt settings laser sim}
\end{figure*}

Results are reported in Fig.~\ref{fig:opt settings laser sim} and Table \ref{tab: q vals sim}. For slow pulse generation, the value of $v^{\pt{2}}$ is almost null for every choice of tunable settings, leading to an estimate of large standard deviation for the conditional probability distribution and, consequently, a value of $q$ close to unity. On the other hand, by increasing the repetition rate we encounter noticeably higher maximum values of $v^{\pt{2}}$, which correspond to less randomness in the pulse phase. This trend is also observed by increasing the value of $I_{off}$, as it implies a larger share of photons from the previous modulation period remains in the cavity.

The same trends are observed also by performing a similar simulation with the assumption $\ell_c =1$, following the scheme reported in Sec. \ref{sec:FirstOrder}. The results are reported in Table \ref{tab: first order q vals sim}. Importantly, we see that for the same laser model an assumption $\ell_c=2$ only slightly worsens the security levels $q$ with respect to the case $\ell_c=1$. This suggests that {the impact of higher order correlations in the scenario simulated is minimal.} Still, to properly prove security one should combine our results in such scenario with the analysis of \cite{pereiraQuantumKey2024}, providing a suitable \textit{ansatz} of the way correlations fade. 

\begin{figure}[ht]
    \centering
    \includegraphics[width=\columnwidth]{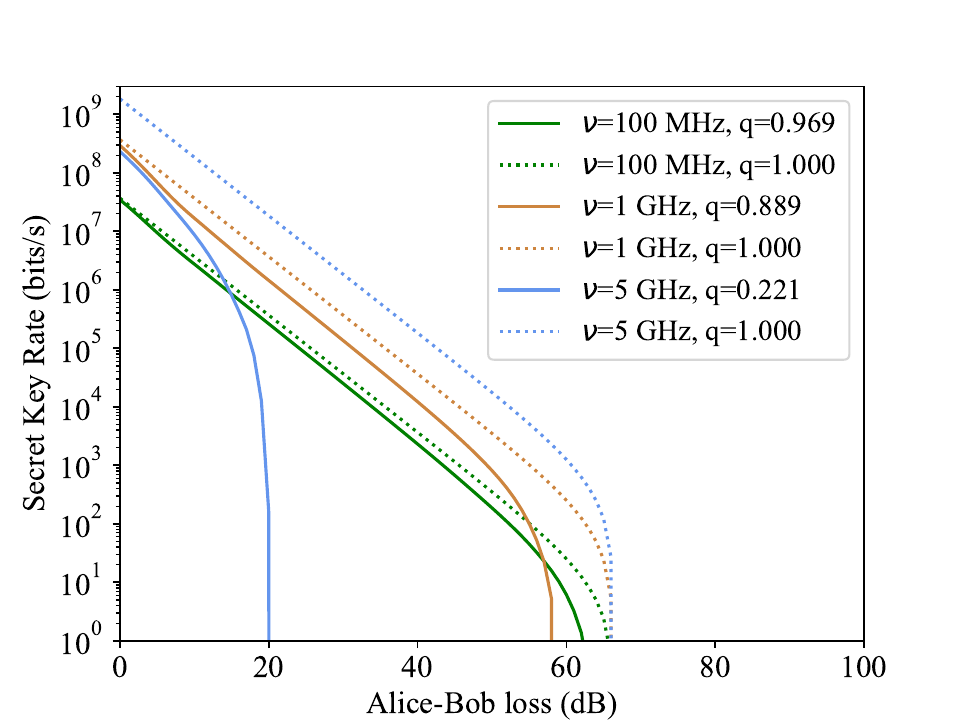}
    \caption{Secret-key rate of the decoy-state BB84 protocol with imperfect phase randomisation (solid lines), compared to the case of perfect phase randomisation (i.e., $q=1$, dotted lines) as a function of the overall system loss for different repetition rates $\nu$ of the source. We consider the results of our simulation from Table \ref{tab: q vals sim} for $I_{off} = 0$ mA and compute key rates through the analysis in \cite{curras-lorenzoSecurityQuantum2024}.
    We assume three intensity settings $\mu_s > \mu_w > \mu_v = 0$. Moreover, for simplicity, we set $\mu_w = \mu_s/5$, and optimise over $\mu_s$. We consider a dark count probability $p_d = 10^{-8}$ for Bob’s detectors, and an error correction inefficiency $f = 1.16$.}
\label{fig:SKR}
\end{figure}

We also note that simulations at $\nu=5$ GHz {differ significantly} from the experiment performed in \cite{grunenfelderPerformanceSecurity2020}, as detailed characterisation of the source parameters for such experiment is not available {and, therefore, our simulation in Fig.~\ref{fig:SKR}} adopts the parameters of the laser in  \cite{rosadoNumericalExperimental2019, quirceSpontaneousEmission2022}.

The secret key rate obtainable for these values of $q$ is shown in Fig.~\ref{fig:SKR}, for the case $I_{off} = 0$ mA. {Here} we use the security proof in \cite{curras-lorenzoSecurityQuantum2024}.
Importantly, these results show that larger values of $q$ allow to {tolerate} higher channel loss. Nevertheless, in case of {low channel loss}, it might be more convenient to employ a source with higher repetition rate but slightly lower value of $q$, since it allows to achieve more secure bits per second. This is the case of a laser source working at $\nu=1$ GHz, compared with the same laser at $\nu=100$ MHz in Fig.~\ref{fig:SKR}.

\begin{table}
        \centering
            \begin{tabular}{|c||*{6}{c|}} \hline
                 $I_{off}$& \multicolumn{2}{|c|}{$0 \ \text{mA}$} & \multicolumn{2}{|c|}{$7 \ \text{mA}$} & \multicolumn{2}{|c|}{$14 \ \text{mA}$} \\
                \hline \hline              
                $\nu $& $\sigma_2$ & $q$ & $\sigma_2$ & $q$ & $\sigma_2$ & $q$ \\
                \hline
                $100 \ \text{MHz}$ &  {$3.238$}&{$0.969$}&  {$3.154$}&{$0.964$} & {$2.972$}&{$0.950$}\\ 
                \hline              
                $1 \ \text{GHz}$ &  {$2.775$}&{$0.889$} &  {$2.999$}&{$0.943$} &  {$3.182$}&{$0.961$} \\    
                \hline
                $ 5 \ \text{GHz}$ &  {$1.730$}&{$0.221$}&  {$0.559$}&{$<0.001$} & {$0.187$}&{$<0.001$}\\
                \hline              
                $ 10 \ \text{GHz}$ &  {$0.061$}&{$<0.001$} &  {$0.050$}&{$<0.001$}&  {$0.041$}&{$<0.001$}\\
                \hline              
            \end{tabular}
        \caption{Numerical results of our simulations in the $\ell_c=2$ scenario for different choices of the laser repetition rate $\nu$ and off-current $I_{off}$. The standard deviation $\sigma_2$ of the phase conditional PDF and value of $q$ are estimated following the scheme introduced in Sec. \ref{sec:SecondOrder}, based on the results in Fig.~\ref{fig:opt settings laser sim}.}
        \label{tab: q vals sim}
    \end{table}   
\begin{table}
        \centering
            \begin{tabular}{|c||*{6}{c|}} \hline
                 $I_{off}$& \multicolumn{2}{|c|}{$0 \ \text{mA}$} & \multicolumn{2}{|c|}{$7 \ \text{mA}$} & \multicolumn{2}{|c|}{$14 \ \text{mA}$} \\
                \hline \hline              
                $\nu $& $\sigma_1$ & $q$ & $\sigma_1$ & $q$ & $\sigma_1$ & $q$ \\
                \hline
                $100 \ \text{MHz}$ &  {$3.362$}&{$0.986$}&  {$3.236$}&{$0.979$} & {$3.095$}&{$0.967$}\\ 
                \hline              
                $1 \ \text{GHz}$ &  {$2.773$}&{$0.916$} &  {$3.000$}&{$0.956$} &  {$3.222$}&{$0.978$} \\    
                \hline
                $ 5 \ \text{GHz}$ &  {$1.728$}&{$0.281$}&  {$0.559$}&{$<0.001$} & {$0.192$}&{$<0.001$}\\
                \hline              
                $ 10 \ \text{GHz}$ &  {$0.061$}&{$<0.001$} &  {$0.071$}&{$<0.001$}&  {$0.110$}&{$<0.001$}\\
                \hline              
            \end{tabular}
        \caption{Numerical results of our simulations in the $\ell_c=1$ scenario for different choices of the laser repetition rate $\nu$ and off-current $I_{off}$. The standard deviation $\sigma_1$ of the phase conditional PDF and value of $q$ are estimated following the scheme reported in Sec. \ref{sec:FirstOrder}.}
        \label{tab: first order q vals sim}
    \end{table} 

\FloatBarrier
\afterpage{\clearpage}

Finally, we remark that in Fig.~\ref{fig:SKR} no secret key rate is displayed for a repetition rate of $\nu=10$ GHz. This is due to the fact that the
analysis in \cite{curras-lorenzoSecurityQuantum2024} does not allow to retrieve positive key rates for $q=0$. Still, it is known that even in the case of WCPs with nonrandom phases, a secret key can be extracted 
\cite{loSecurityQuantum2007}.

\section{Conclusions}\label{sec: Conclusions}
Practical implementations of quantum key distribution (QKD) with laser sources commonly rely on the decoy-state method, whose security is seriously threatened by phase correlations in gain-switched {lasers} driven at high repetition rates \cite{kobayashiEvaluationPhase2014, grunenfelderPerformanceSecurity2020}. In this work, we have tackled this problem and provided experimental techniques to characterise {these correlations}, quantifying their impact on the phase {probability density function} and enabling the application of the security proof in \cite{curras-lorenzoSecurityQuantum2024} for any correlation length $\ell_c$. 

In detail, our design adopts a linear optics network to replicate the behaviour of residual photons from prior modulation periods in the laser cavity. For parameters whose direct measurement is experimentally challenging, we {have proven} that an optimisation task over the settings of the network devices suffices in providing a reliable estimate of the security level of the source. Moreover, we have shown by numerical means the application of our approach to a realistic device in the case of second order correlations.
In this respect, our work significantly narrows the gap between high-speed industrial implementations of QKD systems and their security proofs.

Finally, we underline that this work might find applications beyond QKD, as it provides a tool for the characterisation of phase correlations in any optical setup employing gain-switching lasers.

\section{Acknowledgements}
The authors thank Fadri Gr\"unenfelder, Giuseppe Vallone and Marco Avesani for insightful discussions.
The work of A. Marcomini is fully funded by the Marie Sklodowska-Curie Grant No. 101072637
(Project Quantum-Safe-Internet).
Support from the Galician Regional Government (consolidation of Research Units:
AtlantTIC), the Spanish Ministry of Economy and Competitiveness (MINECO), the
Fondo Europeo de Desarrollo Regional (FEDER) through the grant No.
PID2020-118178RB-C21, MICIN with funding from the European Union
NextGenerationEU (PRTR-C17.I1) and the Galician Regional Government with own
funding through the “Planes Complementarios de I+D+I con las Comunidades
Autónomas” in Quantum Communication, and the European Union’s Horizon Europe
Framework Programme under the project “Quantum Security Networks Partnership”
(QSNP, grant agreement No. 101114043) is greatly acknowledged. A. Valle acknowledges Ministerio de Ciencia e Innovación, PID2021-12345OB-C22 MCIN/AEI/FEDER, UE. K. Tamaki acknowledges support from JSPS KAKENHI Grant {Number} 23H01096.

\appendix
\section{Note on the cavity model}\label{sec: Note on the cavity model}
In this {Appendix} we discuss the underlying reasoning that provides a physical justification for Assumption (A4).

Authors in \cite{kobayashiEvaluationPhase2014} claim that the phase difference between subsequent pulses in gain-switched lasers is a Gaussian variable. Indeed, this argument is consistent with decades of studies on phase diffusion in lasers \cite{laxClassicalNoise1967, henryTheoryLinewidth1982, henryTheoryPhase1983} that led to the formulation of the stochastic rate equations with Langevin forces \cite{septrianiParametricStudy2020, bennettQuantumCryptography1984}. The rationale behind this model comes from observing that the quantum processes building light up inside the cavity affect the field phase in different ways: stimulated emission of photons amplifies the coherent component of light, while spontaneous emissions act as random noise with uniform angular phase distribution (also called ``white" or ``Gaussian" noise). 

More formally, in a single-mode laser the quantum mechanical processes happening in the cavity effectively amplify the residual amount of photons from previous modulation periods leading to an output coherent state at round $i$ \cite{glauberCoherentIncoherent1963}:
    \begin{equation}\label{eqn: Glauber for sum of alpha}
        \ket{\alpha_i} = \ket{\alpha^{cc}_i + \alpha^{sp}_i} := \ket{\sum_{n=i-\ell_c}^{i-1} \alpha_n + \sum_{m=1}^{M_{sp}} e^{j\pmb{\phi}_{m}^{sp}}},
    \end{equation}
    where $\alpha^{cc}_i$ is the complex term denoting the coherent component due to residual photons, while $\alpha^{sp}_i$ represents all the $M_{sp}$ spontaneous emissions. In the definition of $\alpha^{cc}_i$, we have that $\alpha_n = \sqrt{\mu_n}e^{j\pmb{\phi}_{n}}$ for some (potentially unknown) amplitudes $\pg{\sqrt{\mu_n}}_n$, while $\pmb{\phi}_{n}$ is the phase of the $n-$th pulse. As for $\alpha^{sp}_i$, we remark that each spontaneous emission takes the form of coherent states with unitary length and uniformly random phase $\pmb{\phi}_{m}^{sp}$. 
    
    Graphically, Eq.~\eqref{eqn: Glauber for sum of alpha} means that the laser output at round $i$ is a coherent state whose representation $\alpha_i$ in the complex plane is given by the vector sum of the amplified coherent components of $\ell_c$ anterior rounds, together with spontaneous emissions. As the latter are uniformly distributed, all the information about the most probable phase of the $i-$th pulse is given by $\alpha^{cc}_i$. This is indeed what Eq.~\eqref{eqn: phi hat generic} represents, with the clear correspondence $r_n = \sqrt{\mu_n}$, up to a global scaling constant. However, due to spontaneous emissions, the phasor $\alpha^{cc}_i$ undergoes an angular random walk \cite{henryTheoryPhase1983}. For $M_{sp}\gg 1$, the PDF associated to this stochastic process is a Gaussian distribution whose standard deviation is proportional to the ratio between the number of spontaneously emitted photons and {the} total amount of photons in the cavity \cite{loudonQuantumTheory2000}. This concept is summarised by Eq.~\eqref{eqn: introcudion of phi hat and delta phi}.

From a physical perspective, the phase can be treated as an initial time shift of the field. In this respect, it is correct to consider it taking values in $\pt{-\infty,\infty}$. However, when it comes to 
practical measurements it is impossible for Eve to distinguish between phases modulo $2\pi$. Hence, in security proofs the phase is considered as a quantity $\phi \in (-\pi,\pi]$ that is experimentally accessible. 
In this framework, instead of Gaussian PDFs it is meaningful to consider wrapped
Gaussian (WG) distributions defined as
\begin{equation}\label{eqn: wrapped gaussian definition}
    \wg{x;\hat{x},\sigma} := \sum_{k=-\infty}^{\infty} f_{G}\pt{x + 2k\pi; \hat{x}, \sigma},
\end{equation}
with $x \in (-\pi,\pi]$, and where $f_G$ denotes the Gaussian distribution over $\mathbb{R}$ with central value $\hat{x}$ and standard deviation $\sigma$ \cite{curras-lorenzoSecurityQuantum2024, septrianiParametricStudy2020}.
These considerations, together with the ones in the previous paragraphs, lead naturally to the definition of the model in Assumption (A4).

It is important to point out how in this model the central value and dispersion of the distribution give complementary information. The former describes \textit{how} previous phases condition the next one, while the latter tells \textit{how much} they affect it. In this respect, the variance of the distribution is the critical parameter for security. In fact, we note that
\begin{align}
    \lim_{\sigma \to 0} \wg{x;\hat{x},\sigma} &= \delta\pt{x-\hat{x}} , \\
    \lim_{\sigma \to \infty} \wg{x;\hat{x},\sigma} &= \frac{1}{2\pi},
\end{align}
where $\delta \pt{x-\hat{x}}$ denotes the Dirac delta distribution. As a consequence, when $\sigma \gg 1$, spontaneous emissions dominate and all information about the past is lost.

We also remark {that} the standard deviation might be different for various orders $\ell_c$, and therefore {we denote it by $\sigma_{\ell_c}$}.
Here we assume it to be independent on the values of the previous phase realisations, from which follows that the quantities $\sigma_{\ell_c}$ and $\hat{\pmb{\phi}}_i^{\pt{\ell_c}}$ in Eq.~\eqref{eqn: phi hat generic} are uncorrelated. However, this might not always be the case, as the coherent component of light $\alpha^{cc}_i$ in Eq.~\eqref{eqn: Glauber for sum of alpha} could have stronger or weaker intensity depending on the alignment of the previous phases, hence the impact of stimulated emissions may vary.
Nevertheless, it is expected that high-order phase correlations in a realistic scenario only arise for strongly coherent processes that confine the pulse phases within a short range and therefore their interference is quasi-constructive.
This also represents the worst case for security, as the impact of spontaneous emissions is minimal.

The state amplitudes $\pg{\sqrt{\mu_n}}_n$ are also assumed constant through the process. Heuristically, this is equivalent to state that the impact of phases from previous rounds on the present pulse only depends on the total time the photons spent in the laser cavity, and not on the values taken by the phase realisations. 

While $\pg{\sqrt{\mu_n}}_n$ and $\sigma_{\ell_c}$ being constant parameters of the system through the rounds (Assumption (A3)) is a necessary condition for mathematical treatment of the real system, the argument above indicates that this is indeed a reasonable approximation, at least within the coherence time of the source.

\section{Proof of Eq.~\eqref{eqn: A4 assumption model A}}\label{app: proof of PDF assumption}
In this Appendix we derive the result in Eq.~\eqref{eqn: A4 assumption model A}.

Consider Assumption (A4). We remind the reader that $\overline{\delta\phi}$ and $\sigma_{\ell_c}$ denote the mean and standard deviation of the WG distribution of $\pmb{\delta\phi}_i$, respectively. 
We have%
\begin{align}
f(\pmb{\delta\phi}_i = \delta\phi_i \vert \pmb{\phi}_{i-1} = \phi_{i-1}, ..., \pmb{\phi}_{i-\ell_c} &= \phi_{i-\ell_c})  \\
&= f(\pmb{\delta\phi}_i = \delta\phi_i ) 
 \nonumber\\
&= f_{\rm WG} (\delta\phi_i,\overline{\delta\phi},\sigma_{\ell_c}),
\nonumber
\end{align}
as $\pmb{\delta\phi}_i$ is independent of \{$\pmb{\delta\phi}_1,...,\pmb{\delta\phi}_{i-1}$\} {and, therefore,} also independent of \{$\pmb{\phi}_1,...,\pmb{\phi}_{i-1}$\}, since the value of the latter set of random variables is specified by the value of the former set. 

We note that it holds:
\begin{widetext}
\begin{align}
f(\pmb{\phi}_i =\phi_i \vert \pmb{\phi}_{i-1} = \phi_{i-1}, ..., \pmb{\phi}_{i-\ell_c} = \phi_{i-\ell_c}) 
&= f(\hat{\pmb{\phi}}_i^{(\ell_c)} + \pmb{\delta\phi}_i =\phi_i  \vert \pmb{\phi}_{i-1} = \phi_{i-1}, ..., \pmb{\phi}_{i-\ell_c} = \phi_{i-\ell_c}) \nonumber \\
&= f(\pmb{\delta\phi}_i =\phi_i -  \hat{\pmb{\phi}}_i^{(\ell_c)} \vert \pmb{\phi}_{i-1} = \phi_{i-1}, ..., \pmb{\phi}_{i-\ell_c} = \phi_{i-\ell_c}) \nonumber \\
&= f(\pmb{\delta\phi}_i =\phi_i - h(\pmb{\phi}_{i-\ell_c},...,\pmb{\phi}_{i-1})  \vert \pmb{\phi}_{i-1} = \phi_{i-1}, ..., \pmb{\phi}_{i-\ell_c} = \phi_{i-\ell_c}) \nonumber \\
&= f(\pmb{\delta\phi}_i = \phi_i - h(\phi_{i-\ell_c},...,\phi_{i-1}) \vert \pmb{\phi}_{i-1} = \phi_{i-1}, ..., \pmb{\phi}_{i-\ell_c} = \phi_{i-\ell_c}) \nonumber \\
&= f(\pmb{\delta\phi}_i = \phi_i - \hat{\phi}_i^{(\ell_c)} \vert \pmb{\phi}_{i-1} = \phi_{i-1}, ..., \pmb{\phi}_{i-\ell_c} = \phi_{i-\ell_c}) \nonumber \\
&= f_{\rm WG} (\phi_i - \hat{\phi}_i^{(\ell_c)},\overline{\delta\phi},\sigma_{\ell_c})
=  f_{\rm WG} (\phi_i,\overline{\delta\phi} + \hat{\phi}_i^{(\ell_c)},\sigma_{\ell_c}),
\end{align}
\end{widetext}
where we have defined $\hat{\phi}_i^{(\ell_c)} = h(\phi_{i-\ell_c},...,\phi_{i-1})$. 

\noindent This concludes the proof.

\section{Derivation of generalised visibilities}\label{AppSec: g}

In this {Appendix} we derive the full expressions of the generalised visibility measures in Eqs.~\eqref{eqn: estimate for sigma_2 in expscheme}-\eqref{eqn: definitioon of g for ell_c}.

\subsection{Second order - Eq.~\eqref{eqn: estimate for sigma_2 in expscheme}}

Consider Fig.~\ref{fig:exp_setup} for reference.
Allowing for an attenuator module $A$ in the longer arm and a local phase shift $\varphi$ in the shorter arm, the three circulating states at round $i$ in the interferometer {are given by Eq.~\eqref{eqn: alpha prime states main text}}.

Let us focus on the recombination of the beams. The states $\ket{\xi}$ and $\ket{\chi}$ take the form 
\begin{equation}\label{eqn: action of BS for ellc=2}
\begin{gathered}
    \ket{\xi} = \ket{\frac{\alpha'_{i-1}-\alpha'_{i-2}}{\sqrt{2}}}, \\
    \ket{\chi} = \ket{\frac{\alpha'_{i-1}+\alpha'_{i-2}}{\sqrt{2}}}. 
\end{gathered}
\end{equation}
This means that $\chi \equiv \sqrt{\pmb{\mu}_{\chi}}e^{j\pmb{\phi}_{\chi}}$, with
\begin{align}\label{eqn: chi definition 1}
    \pmb{\mu}_{\chi} &= \frac{1}{2}\pt{\mu'_{i-1} + \mu'_{i-2}} \nonumber \\ 
    &  + \sqrt{\mu'_{i-1}\mu'_{i-2}}\cos\pt{\pmb{\phi}_{i-2} - \pmb{\phi}_{i-1}} ,\\
    \pmb{\phi}_{\chi} &= \arg\pt{e^{j\pmb{\phi}_{i-1}} + \sqrt{\frac{\mu'_{i-2}}{\mu'_{i-1}}}e^{j\pmb{\phi}_{i-2}}} \nonumber \\
     &=\arg\pt{e^{j\pmb{\phi}_{i-1}} + r'_{i-2} e^{j\pmb{\phi}_{i-2}}}.
     \label{eqn: chi definition 2}
\end{align}

Direct comparison with Eq.~\eqref{eqn: phi hat generic} in the case $\ell_c=2$ shows that when $r'_{i-2} = r_{i-2}$ the phase of $\ket{\chi}$ is equal to $\hat{\pmb{\phi}}_i^{\pt{2}}$, at every round.
Nevertheless, satisfying this condition is very challenging in a realistic case. Due to the ignorance of the exact value of $r_{i-2}$ and to the imperfect tuning of the amplitudes, the phase $\pmb{\phi}_{\chi}$ might be shifted from $\hat{\pmb{\phi}}_i^{\pt{2}}$ by a small quantity $\pmb{\epsilon}'_i$. Note that $\pmb{\epsilon}'_i$ depends on the {pulses' amplitudes} and phases themselves, and accounts for any systematic error in the optical path length. In fact, let us define
\begin{eqnarray}
\delta r \ &&:= r'_{i-2} - r_{i-2} ,\\
\hat{\alpha}_i^{\pt{2}} \ &&:= e^{j\pmb{\phi}_{i-1}} + r_{i-2} e^{j\pmb{\phi}_{i-2}},
\end{eqnarray}
such that $\arg\pt{\hat{\alpha}_i^{\pt{2}}} = \hat{\pmb{\phi}}_i^{\pt{2}}$. We have (modulo $2\pi$):
\begin{eqnarray}
    \pmb{\phi}_{\chi} &&= \arg\pt{e^{j\pmb{\phi}_{i-1}} + r'_{i-2} e^{j\pmb{\phi}_{i-2}}} \nonumber \\
    &&= \arg\pt{\hat{\alpha}_i^{\pt{2}} + \delta r e^{j\pmb{\phi}_{i-2}}} \nonumber \\
    &&= \hat{\pmb{\phi}}_i^{\pt{2}} + \arg\pt{1+\frac{\delta r}{\abs{\hat{\alpha}_i^{\pt{2}}}}e^{j\pt{\pmb{\phi}_{i-2}-\hat{\pmb{\phi}}_i^{\pt{2}}}}} \nonumber \\
    && =: \hat{\pmb{\phi}}_i^{\pt{2}} + \pmb{\epsilon}'_i .
    \label{eqn: definition of epsilon'_i}
\end{eqnarray}
Importantly, while $\pmb{\epsilon}'_i$ depends on the actual values of the phases of previous rounds, it goes to zero as $r'_{i-2}$ approaches $r_{i-2}$.

By Assumption (A4), the random process associated to the set of variables $\pg{\pmb{\delta\phi}_i}_i$ consists of a set of identically distributed Gaussian variables following a distribution
\begin{equation}
    f(\pmb{\delta\phi}_i = \delta\phi_i) = 
    \wg{\delta\phi_i;\overline{\delta\phi},\sigma_2}.
\end{equation}
One can easily find that
\begin{equation}
    \expval{\exp\pt{j \pmb{\delta\phi}_i}} = \exp\pt{-\frac{\sigma_2^2}{2}} e^{j\overline{\delta\phi}} ,
\end{equation}
where $\expval{X}$ denotes the ensemble average of the random variable $X$. Therefore
\begin{equation}\label{eqn: sin for deltaphi_i}
    \expval{\sin\pt{\pmb{\delta\phi}_i}} = \exp\pt{-\frac{\sigma_2^2}{2}}\sin\pt{\overline{\delta\phi}} ,
\end{equation}
\begin{equation}\label{eqn: cos for deltaphi_i}
    \expval{\cos\pt{\pmb{\delta\phi}_i}} = \exp\pt{-\frac{\sigma_2^2}{2}}\cos\pt{\overline{\delta\phi}} .
\end{equation}

At the final BS in Fig.~\ref{fig:exp_setup}, the input states are $\ket{\alpha'_i}$ and $\ket{\chi}$. 
Assuming lossless BSs for simplicity, conservation of energy implies that an alternative expression for $\pmb{\mu}_{\chi}$ is given by
\begin{equation}\label{eqn: conservation of energy BS for mu chi}
    \pmb{\mu}_{\chi} = \mu'_{i-1} + \mu'_{i-2} - \pmb{\mu}_{\xi} ,
\end{equation} 
where $\pmb{\mu}_{\xi}$ represents the intensity of the state $\ket{\xi}$.
The output states entering the detectors have the following form
\begin{equation}
\begin{gathered}
    \ket{\beta_1}= \ket{\frac{\alpha'_{i}+\chi}{\sqrt{2}}} ,
     \\
     \ket{\beta_2} = \ket{\frac{\alpha'_{i}-\chi}{\sqrt{2}}} ,
\end{gathered}
\end{equation}  
and the measured energies are proportional to
\begin{eqnarray}\label{eqn: I_12}
    \pmb{\mu}_{\beta_{1,2}} &&= \frac{1}{2}\pq{\abs{\alpha'_i}^2 + \abs{\chi}^2 \pm \pt{{\alpha'_i}^*{\chi}+  {\alpha'_i}\chi^*}} \\
    &&= \frac{1}{2}\pt{\mu'_i + \pmb{\mu}_{\chi}} \pm \sqrt{\mu'_i \pmb{\mu}_{\chi}}\cos\pt{\pmb{\phi}_i - \pmb{\phi}_{\chi} + \varphi} \nonumber \\
    &&= \frac{1}{2}\pt{\mu'_i + \pmb{\mu}_{\chi}} \pm \sqrt{\mu'_i \pmb{\mu}_{\chi}}\cos\pt{\pmb{\phi}_i - \hat{\pmb{\phi}}_i^{\pt{2}} - \pmb{\epsilon}'_i + \varphi} \nonumber \\
    &&= \frac{1}{2}\pt{\mu'_i + \pmb{\mu}_{\chi}} \pm \sqrt{\mu'_i \pmb{\mu}_{\chi}}\cos\pt{\pmb{\delta\phi}_i - \pmb{\epsilon}'_i + \varphi},
    \nonumber
\end{eqnarray}
which implies

\begin{equation}\label{eqn: simple b1b2 for cos()}
    \cos\pt{\pmb{\delta\phi}_i - \pmb{\epsilon}'_i + \varphi} = \frac{\pmb{\mu}_{\beta_1} - \pmb{\mu}_{\beta_2}}{2\sqrt{\mu'_i\pmb{\mu}_{\chi}}}.
\end{equation}

Assuming again lossless BSs, we have
\begin{equation}
    \pmb{\mu}_{\beta_2} = \mu'_i + \pmb{\mu}_{\chi} - \pmb{\mu}_{\beta_1}.
\end{equation}
Substituting in Eq.~\eqref{eqn: simple b1b2 for cos()},
an equivalent form explicitly dependent on the attenuated intensity $\mu'_{i-2}$ is given by
\begin{equation}\label{eqn: cos for v2 def}
    \cos\pt{\pmb{\delta\phi}_i - \pmb{\epsilon}'_i + \varphi} \ %
    = \frac{{2\pmb{\mu}_{\beta_1} + \pmb{\mu}_{\xi}} - \pt{\mu'_i + \mu'_{i-1} + \mu'_{i-2}}}{2\sqrt{\mu'_i\pt{\mu'_{i-1} + \mu'_{i-2} - \pmb{\mu}_{\xi}}}}. 
\end{equation}
Note that this expression is more convenient, as it does not require to monitor the output state $\ket{\beta_2}$.

Taking the ensemble average of Eq.~\eqref{eqn: cos for v2 def} yields the quantity {$v^{\pt{2}}$} defined in Eq.~\eqref{eqn: definitioon of g}. Moreover, we have that
\begin{align} \nonumber
    v^{\pt{2}} := &\expval{\cos\pt{\pmb{\delta\phi}_i - \pmb{\epsilon}'_i + \varphi}} \nonumber \\
    = &\expval{\cos\pt{\pmb{\delta\phi}_i + \varphi}\cos\pt{\pmb{\epsilon}'_i} + \sin\pt{\pmb{\delta\phi}_i + \varphi}\sin\pt{\pmb{\epsilon}'_i}} \nonumber \\
    = &\exp\pt{-\frac{\sigma_2^2}{2}}\cos\pt{\overline{\delta\phi}+\varphi}\expval{\cos\pt{\pmb{\epsilon}'_i}} \nonumber \\
    &+ \exp\pt{-\frac{\sigma_2^2}{2}}\sin\pt{\overline{\delta\phi}+\varphi}\expval{\sin\pt{\pmb{\epsilon}'_i}} ,
    \label{eqn: long and full for expval v2}
\end{align}
where the last equality holds since, by Assumption (A4), $\pmb{\delta\phi}_i$ is independent of $\pmb{\phi}_{i-1}$ and $\pmb{\phi}_{i-2}$ (and thus of $\pmb{\epsilon}'_i$).

As argued, as $r'_{i-2} \to r_{i-2}$ we observe $\pmb{\epsilon}'_i \to 0$, which ultimately implies $\expval{\cos\pt{\pmb{\epsilon}'_i}} \to 1$ and $\expval{\sin\pt{\pmb{\epsilon}'_i}} \to 0$ for every round $i$.
Therefore, the maximum value of Eq.~\eqref{eqn: long and full for expval v2} is found for the optimal settings choices we reported in Eqs.~\eqref{eqn: optimal paramteres main text 1}-\eqref{eqn: optimal paramteres main text 2}. Thus, we obtain that
\begin{equation}
   \max_{\varphi,A} v^{\pt{2}} = \exp\pt{-\frac{\sigma_2^2}{2}} ,
\end{equation}
from which Eq.~\eqref{eqn: estimate for sigma_2 in expscheme} follows naturally.

Finally, let us elaborate on an important note about the concept of visibility. The visibility measure introduced in Eq.~\eqref{eqn: visibility definition} depends on the light intensity $I$ considered as an ensemble average of the detector output for fixed experimental settings \cite{kobayashiEvaluationPhase2014}. In our case, for fixed $\varphi$ and $A$, one would find an average intensity
\begin{equation}\label{eqn: I generic for phi and A}
    I\pt{\varphi,A} = \frac{1}{2}\pt{{\mu'_i} + \expval{\pmb{\mu}_{\chi}}} + \expval{\sqrt{\mu'_i \pmb{\mu}_{\chi}}\cos\pt{\pmb{\phi}_i - \pmb{\phi}_{\chi} + \varphi}},
\end{equation}
for the detector measuring $\ket{\beta_1}$. Note that here $\expval{\mu'_{i}} = \mu'_i$ as this intensity is fixed by the interferometer architecture.
Crucially, for $\ell_c = 2$, $\pmb{\mu}_{\chi}$ and $\pmb{\phi}_{\chi}$ are not statistically independent quantities, as they both have direct dependence on $\pmb{\phi}_{i-1}$ and $\pmb{\phi}_{i-2}$ (see Eqs.~\eqref{eqn: chi definition 1}-\eqref{eqn: chi definition 2}). As a consequence, for the generic case $\ell_c \ge 2$, we have that
\begin{align}
    &\expval{\sqrt{\mu'_i \pmb{\mu}_{\chi}}\cos\pt{\pmb{\phi}_i - \pmb{\phi}_{\chi} + \varphi}} \nonumber \\
    &\quad \quad \quad \neq \sqrt{{\mu'_i} \expval{\pmb{\mu}_{\chi}}}\expval{\cos\pt{\pmb{\phi}_i - \pmb{\phi}_{\chi} + \varphi}}.
    \label{eqn: vissibility not equal expval}
\end{align}
Therefore, proving that a standard visibility measure could allow to estimate $\sigma_2$ is not straightforward, as one would need further assumptions to be able to relate $\sigma_2$ to average intensities in the form of Eq.~\eqref{eqn: I generic for phi and A}.
Nevertheless, by rescaling the output signal at each round by a factor $\sqrt{\mu'_i \pmb{\mu}_{\chi}}$ and then taking the average, we can introduce an alternative metric that focuses on the interference term and decouples $\pmb{\mu}_{\chi}$ and $\pmb{\phi}_{\chi}$, that is, Eq.~\eqref{eqn: long and full for expval v2}. 
Importantly, as we show below, our definition converges to the canonical visibility measure in the case $\ell_c=1$, but can be fully applied to arbitrary correlation length. Therefore, we shall refer to this quantity as ``generalised" (or ``higher order") visibility.

Consider now the case $\ell_c=1$. By removing the longer arm of the interferometer in Fig.~\ref{fig:exp_setup}, we have 
\begin{equation}
    \pmb{\mu}_{\chi} = \mu'_{i-1}, \quad \pmb{\phi}_{\chi} = \pmb{\phi}_{i-1} = \hat{\pmb{\phi}}^{\pt{1}}_i, 
\end{equation}
the former value being now constant from Assumption (A1). Also, we removed the amplitude attenuator $A$ as now we have $\delta r = 0 \implies \pmb{\epsilon}'_i = 0$ in Eq.~\eqref{eqn: definition of epsilon'_i}.
Therefore, the argument of Eq.~\eqref{eqn: vissibility not equal expval} does not hold for this case and by means of Eq.~\eqref{eqn: I generic for phi and A}, one finds
\begin{equation}\label{eqn:I lc_1}
    I\pt{\varphi} = \frac{1}{2}\pt{{\mu'_i} + {\mu'_{i-1}}} + \sqrt{\mu'_i \mu'_{i-1}}\expval{\cos\pt{\pmb{\delta\phi}_i + \varphi}}.
\end{equation}
Now, it follows from Eq.~\eqref{eqn:I lc_1} that the standard visibility introduced in Eq.~\eqref{eqn: visibility definition} reads
\begin{equation}\label{eqn: visibility of lc_1}
    \mathcal{V} = \frac{2\sqrt{\mu'_i \mu'_{i-1}}\expval{\cos\pt{\pmb{\delta\phi}_i + \varphi_{max}}}}{{\mu'_i} + {\mu'_{i-1}}},
\end{equation}
and since we are adopting 50:50 BSs we ultimately have $\mu'_i = \mu'_{i-1} = \mu$ and, therefore, $\mathcal{V} = \expval{\cos\pt{\pmb{\delta\phi}_i + \varphi_{max}}}$.
By comparing this last expression with the first line of Eq.~\eqref{eqn: long and full for expval v2} when maximising over $\varphi$,
we conclude that 
\begin{equation}
    {v^{\pt{1}}} \equiv \mathcal{V} ,
\end{equation}
that is, our approach converges to the standard visibility measure for first order correlations.

\subsection{Generalisation to arbitrary order - Eq.~\eqref{eqn: definitioon of g for ell_c}}\label{Appendix: generalisation of g}
Here we generalise the approach for the second order case to correlations of arbitrary {finite} length $\ell_c$.
Again, the aim is to make each pulse interfere with a combination of previous pulses whose phase matches $\hat{\pmb{\phi}}_i^{\pt{\ell_c}}$. 

Consider the general setup shown in Fig.~\ref{fig:exp_setup_inf}. Let \begin{equation}
    \ket{\alpha'_n} = \ket{\sqrt{\mu_{n}'}e^{j\pmb{\phi}_{n}}}
\end{equation}
for $i-\ell_c \le n \le i-1$ be the coherent states in the delay lines in the grey box after amplitude attenuation. At the top right beam splitter the two less recent states $\ket{\alpha'_{i-\ell_c}}$ and $\ket{\alpha'_{i-\ell_c+1}}$ lead to the output states
\begin{equation}
\begin{gathered}
     \ket{\xi_{i-\ell_c+2}} 
     = \ket{\frac{\alpha'_{i-\ell_c+1}-\alpha'_{i-\ell_c}}{\sqrt{2}}},\\
     \ket{\chi_{i-\ell_c+2}} = \ket{\frac{\alpha'_{i-\ell_c+1}+\alpha'_{i-\ell_c}}{\sqrt{2}}},
 \end{gathered}
\end{equation}
and for any other BS in the right hand side we have that inputs $\ket{\alpha'_{n}}$ and $\ket{\chi_{n}}$ are mapped to outputs
\begin{equation}
\begin{gathered}
    \ket{\xi_{n+1}} = \ket{\frac{\alpha'_{n}-\chi_n}{\sqrt{2}}},
    \\
   \ket{\chi_{n+1}} = \ket{\frac{\alpha'_{n}+\chi_n}{\sqrt{2}}}.
\end{gathered}
\end{equation}

Proceeding iteratively, it can be shown that
\begin{equation}\label{eqn: full form chi_i}
    \ket{\chi_i} = \ket{\frac{\alpha'_{i-\ell_c}}{2^{\pt{\ell_c-1}/2}} + \sum_{k=1}^{\ell_c-1} \frac{\alpha'_{i-\ell_c + k}}{2^{\pt{\ell_c-k}/2}} } ,
\end{equation}
which is the generalisation of Eq.~\eqref{eqn: action of BS for ellc=2}.
The scaling factors of the terms in Eq.~\eqref{eqn: full form chi_i} are due to the amount of balanced BSs each state encounters in its optical path and can be taken into account while tuning the attenuations. Hence, for simplicity we will consider them included in the definition of the state amplitudes $\pg{\sqrt{\mu'_n}}$ from now on. 

Following the same idea of the previous section, let
\begin{equation}
    \delta r_{n} := r_{n}' - r_{n} \quad , \quad r_n' := \sqrt{\frac{\mu_n'}{\mu'_{i-1}}} 
\end{equation}
for $i-\ell_c \le n \le i-1$, and
\begin{equation}
    \hat{\alpha}_i^{\pt{\ell_c}} := \sum_{n=i-\ell_c}^{i-1} r_n e^{j\pmb{\phi}_{n}} ,
\end{equation}
so that $\arg\pt{\hat{\alpha}_i^{\pt{\ell_c}}} = \hat{\pmb{\phi}}_i^{\pt{\ell_c}}$. Then, we have that
\begin{eqnarray}
    \pmb{\phi}_{\chi_i} &&= \arg\pt{\sum_{n=i-\ell_c}^{i-1} r_n' e^{j\pmb{\phi}_{n}}} \nonumber \\
    &&= \arg\pt{\hat{\alpha}_i^{\pt{\ell_c}} + \sum_{n=i-\ell_c}^{i-1} \delta r_n e^{j\pmb{\phi}_{n}}} \nonumber \\
    &&= \hat{\pmb{\phi}}_i^{\pt{\ell_c}} + \arg\pt{1+\sum_{n=i-\ell_c}^{i-1} \frac{\delta r_n}{\abs{\hat{\alpha}_i^{\pt{\ell_c}}}} e^{j\pt{\pmb{\phi}_{n} - \hat{\pmb{\phi}}_i^{\pt{\ell_c}}}}} \nonumber \\
    && =: \hat{\pmb{\phi}}_i^{\pt{\ell_c}} + \pmb{\epsilon}'_i.
\end{eqnarray}
Again, we find that $\pmb{\epsilon}'_i \to 0$ as $\delta r_n \to 0 \ \forall n$. 

Let $\pg{\pmb{\mu}_{\xi_k}}_{k=i-\ell_c+2}^{i}$ be the set of intensities of the states $\pg{\ket{\xi_k}}_k$ in Fig.~\ref{fig:exp_setup_inf}.
For lossless BSs, conservation of energy implies that
\begin{equation}
    \pmb{\mu}_{\chi_i} = \sum_{n=i-\ell_c}^{i-1} \mu_n' - \sum_{k=i-\ell_c+2}^{i} \pmb{\mu}_{\xi_k}.
\end{equation}

From here on, the whole reasoning of section \ref{AppSec: g} holds with the natural generalisation $\chi \to \chi_i$.
In particular:
\begin{eqnarray}
    \cos\pt{\pmb{\delta\phi}_i - \pmb{\epsilon}'_i + \varphi} &&= \frac{\pmb{\mu}_{\beta_1} - \pmb{\mu}_{\beta_2}}{2 \sqrt{\mu'_i\mu_{\chi_i}}} \\
    &&= \frac{2 \pmb{\mu}_{\beta_1} + \sum_k\pmb{\mu}_{\xi_k} - \sum_{n=i-\ell_c}^{i}\mu'_n}{2\sqrt{\mu'_i\pt{\sum_{n=i-\ell_c}^{i-1}\mu'_n-\sum_k \pmb{\mu}_{\xi_k}}}} , \nonumber
\end{eqnarray}
whose average corresponds to the value of $v^{\pt{\ell_c}}$ in Eq.~\eqref{eqn: definitioon of g for ell_c}.

\section{Exponential decay of correlations}\label{sec: exponential decay}

\begin{figure}
    \centering
    \includegraphics[width=\columnwidth]{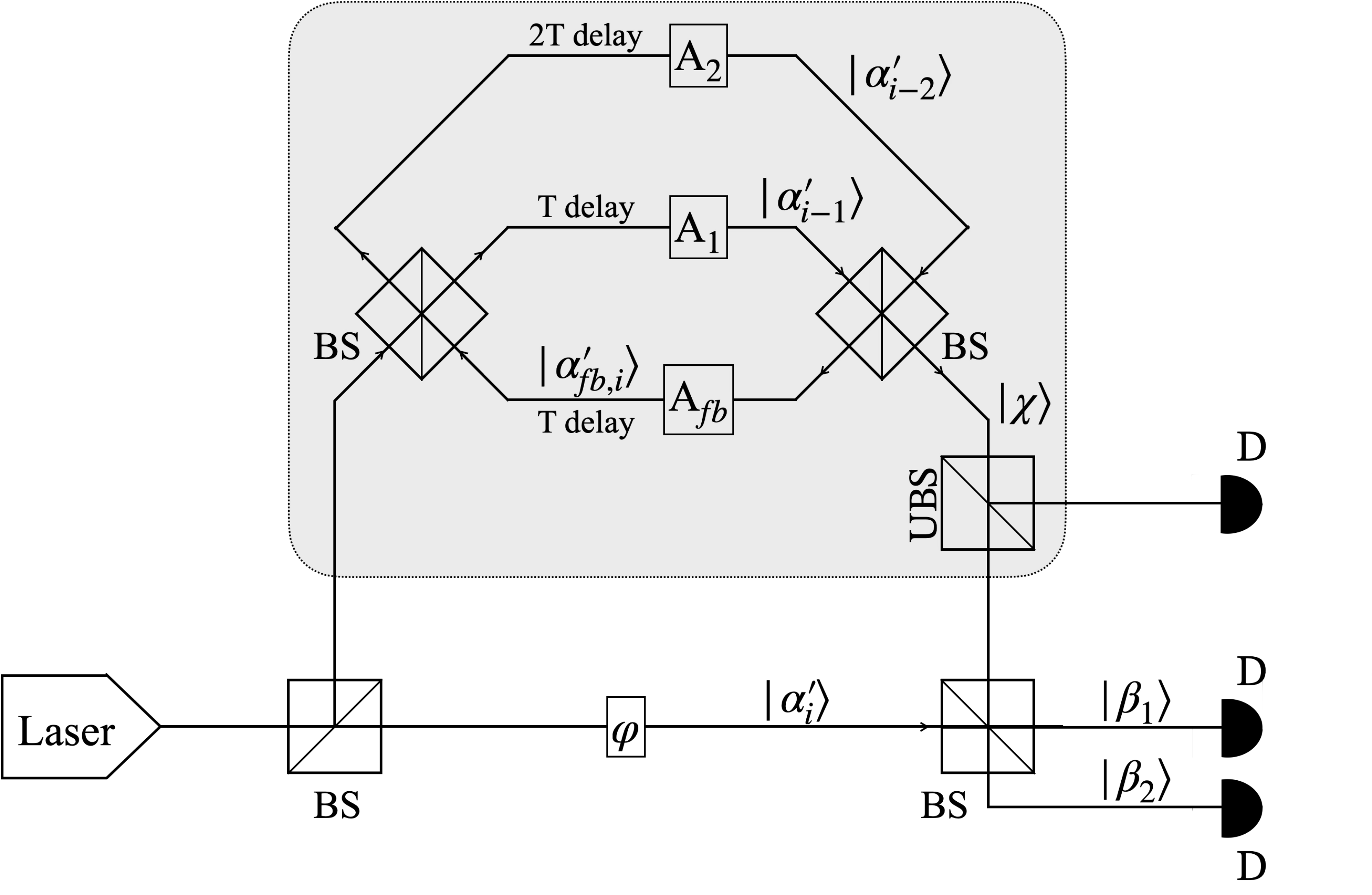}
    \caption{Feedback-loop interferometer scheme for the estimate of quantum noise in the case of correlations with exponential decay. The introduction of the feedback line allows for a share of light to remain in the interferometer at each round, while attenuators periodically dump the intensity of the pulse. Tuning the intensity allows to reproduce the expected value of the next phase $\hat{\pmb{\phi}}_i^{\pt{\ell_c}}$ for $\ell_c\to\infty$ by effectively finding $r_0$ in Eq. \eqref{eqn: r for MZinf}
    {Here, $\varphi$, phase shifter; D, linear detector; $A_k$, attenuators; UBS, unbalanced beam splitter (e.g., 90:10, only required to learn the intensity of the state $\ket{\chi}$). All other beam splitters (BSs) are considered $50:50$.}}
    \label{fig:exp_setup_fb}
\end{figure}

In this section we investigate the particular scenario of exponentially decaying correlations. This is of special interest as both ordinary laser rate equations (e.g., the ones in Eq. \eqref{eqn: diff for N}) and recent experimental results \cite{agulleiroModelingCharacterization2024} suggest that the number of photons in the cavity might deplete exponentially when the laser is below threshold. Therefore, a suitable \textit{ansatz} is that the past state intensities $\pg{\sqrt{\mu_n}}_n$ in Eq. \eqref{eqn: Glauber for sum of alpha} also share the same scaling and therefore
\begin{equation}\label{eqn: r for MZinf}
    r_{i-k} = \sqrt{\frac{\mu_{i-k}}{\mu_{i-1}}} = r_0^{k-1},
\end{equation}
for some $0 < r_0 \le 1 $.
In this case, characterising the conditional phase distribution over the past simply reduces to finding $r_0$, $\overline{\delta\phi}$ and $\sigma_{\ell_c}$. This can be done by means of a feedback-loop interferometer which allows a share of the light at each round to keep interfering with future pulses. A scheme for such setup, derived directly from the one in Fig. \ref{fig:exp_setup} with the addition of a feedback line, is illustrated in Fig. \ref{fig:exp_setup_fb}. Specifically, as for the case $\ell_c=2$ investigated in Sec. \ref{sec:SecondOrder}, the light arriving from the source is split into three channels inducing different delays by means of BSs. As in Fig. \ref{fig:exp_setup}, the states $\ket{\alpha'_{i-2}}$ and $\ket{\alpha'_{i-1}}$ in the two longest arms recombine at another BS. However, in Fig. \ref{fig:exp_setup_fb} one of the outputs of such BS is redirected towards the inputs (state $\ket{\alpha'_{fb,i}}$) on a delay line of length $T$, so to create two loops in the interferometer.
As a result, at round $i$ a fraction of the $\pt{i-n}-$th pulse has undergone through attenuation a number of times proportional to $n$, which implies that it reaches the final BS with exponentially dumped intensity. Hence, the optimisation of a small set of attenuators is enough to replicate the full light build-up in the laser cavity. 

Importantly as well, note that this scheme allows, in principle, for the analysis of correlations of unbounded length that decay exponentially \cite{pereiraQuantumKey2024}.

\section{Simulation details}\label{appsec: LaserSim}
Gain-switched single-mode semiconductor laser dynamics can be modelled by using a set of
stochastic rate-equations that read (in Ito’s sense) \cite{coldren2012diode,agrawal2013semiconductor,valleDivergenceVariance2023,valleStatisticsOptical2021,lovicCharacterizingPhase2021,shakhovoy2023phase}:
\begin{widetext}
\begin{align}
    \dv{E\pt{t}}{t} &= \pq{\pt{\frac{1}{1+\epsilon\abs{E\pt{t}}^2}+j\alpha}G_N \pt{N\pt{t}-N_t} - \frac{1+j\alpha}{\tau_p}}\frac{E\pt{t}}{2} +\sqrt{\beta B} N\pt{t} \xi\pt{t}, \nonumber
    \\
    \dv{N\pt{t}}{t} &= \frac{I\pt{t}}{e} - \pt{AN\pt{t} + BN^2\pt{t} + CN^3\pt{t}} - \frac{G_N\pt{N\pt{t}- N_t}\abs{E\pt{t}}^2}{1+\epsilon\abs{E\pt{t}}^2}
    \nonumber
    \\
    &+\sqrt{2\pt{AN\pt{t} + BN^2\pt{t} + CN^3\pt{t} + \frac{I\pt{t}}{e}}}\xi_N\pt{t} -2\sqrt{\beta B}N\pt{t}\pt{E_1\pt{t}\xi_1\pt{t} + E_2\pt{t}\xi_2\pt{t}}.
    \label{eqn: diff for N}
\end{align}
\end{widetext}
Here, $E\pt{t} = E_1\pt{t} + jE_2\pt{t}$ is the complex electrical field, and $N\pt{t}$ is the number of carriers
in the active region. The term $\xi\pt{t} = \xi_1\pt{t} + j \xi_2\pt{t}$ is the complex Gaussian white noise with zero
average and correlation given by $\expval{\xi\pt{t}\xi^*\pt{t'}} = \delta\pt{t-t'}$ that represents the spontaneous emission noise. The quantity $\xi_N\pt{t}$ is a real Gaussian white noise of zero average and correlation given by $\expval{\xi_N\pt{t}\xi_N\pt{t'}} = \delta\pt{t-t'}$, statistically independent of $\xi\pt{t}$ \cite{fatadin2006numerical, mcdaniel2018stochastic, rosadoNumericalExperimental2019}. 
The parameters appearing in these equations are the following: $G_N$ is the differential gain, $N_{t}$ is the number of carriers at transparency, $\epsilon$ is the non-linear gain coefficient, $\tau_p$ is the photon
lifetime, $\beta$ is the fraction of spontaneous emission coupled into the lasing mode, $\alpha$ is the linewidth
enhancement factor, $e$ is the electron charge, and $A$, $B$ and $C$ are the non-radiative, spontaneous, and
Auger recombination coefficients, respectively. 
At any point in time, the phase of the electric field is given by \begin{equation}
    \phi\pt{t} = \atan\pt{E_2\pt{t}/E_1\pt{t}}.
\end{equation}

To generate pulses, we consider an injected current, $I\pt{t}$, which follows a
periodic square-wave modulation of period $T=1/\nu$ with $I\pt{t} = I_{on} > I_{th}$ during $T/2$ and $I\pt{t} = I_{off} < I_{th}$ during
the rest of the period. Here $\nu$ is the repetition rate of the laser and $I_{th}$ the threshold current. We use the numerical values of the parameters that have been extracted for
a discrete mode laser (DML) \cite{rosadoNumericalExperimental2019, quirceSpontaneousEmission2022} and adopt them to generate an electric field through Eq.~\eqref{eqn: diff for N}, 
which is numerically solved by using the Euler-Maruyama algorithm \cite{riskenFokkerPlanckEquation1996a,kloedenStochasticDifferential1992} with an integration time step of $0.01$ ps. The value of parameters adopted are reported in Table \ref{tab: parameters for sim}.

The threshold current value $I_{th} = 14.14$ mA is obtained from  
\begin{equation}
    I_{th} = e\pt{AN_{th} + BN_{th}^2 + CN_{th}^3},
\end{equation}
where 
\begin{equation}
    N_{th} = N_t + 1/\pt{G_N\tau_p} 
\end{equation}
is the number of carriers at threshold \cite{quircePhaseDiffusion2021}. By setting $I_{on} = 140$ mA, we study the security of the laser when changing the repetition rate $\nu \in \pg{100\ \text{MHz}, 1 \ \text{GHz}, 5 \ \text{GHz}, 10 \ \text{GHz}}$ and setting the current $I_{off}$ far below threshold ($0$ mA, laser off), to an intermediate level ($7$ mA), and slightly below threshold ($14$ mA).

\begin{table}[h!]
\centering
\begin{tabular}{|c|c|}
\hline
\textbf{Parameter} & \textbf{Value} \\ \hline
$G_N$ & $1.48 \times 10^4 \ \text{s}^{-1}$ \\ \hline
$N_t$ & $1.93 \times 10^7$ \\ \hline
$\epsilon$ & $7.73 \times 10^{-8}$ \\ \hline
$\tau_p$ & $2.17 \ \text{ps}$ \\ \hline
$\alpha$ & $3$ \\ \hline
$\beta$ & $5.3 \times 10^{-6}$ \\ \hline
$A$ & $2.8 \times 10^{8} \ \text{s}^{-1}$ \\ \hline
$B$ & $9.8 \ s^{-1}$ \\ \hline
$C$ & $3.84 \times 10^{-7} \ \text{s}^{-1}$ \\ \hline
\end{tabular}
\caption{List of parameters and their values adopted for the laser simulation (from  \cite{rosadoNumericalExperimental2019, quirceSpontaneousEmission2022}).}
\label{tab: parameters for sim}
\end{table}

\bibliographystyle{apsrev4-2.bst}
\bibliography{MyPaper3}%

\end{document}